\documentclass[twocolumn]{aastex631}

\usepackage{amsmath}
\usepackage{hyperref}

\begin{document}

\title{A possible high-redshift origin for the short GRB 061201:\\
implications of a compact binary merger beyond cosmic noon }

\author[0000-0002-1869-7817]{Troja E.}
\affiliation{Department of Physics, University of Rome ``Tor Vergata'', via della Ricerca Scientifica 1, I-00133 Rome, Italy}
\author[0000-0002-9700-0036]{O'Connor, B.}
\affiliation{McWilliams Center for Cosmology and Astrophysics, Department of Physics, Carnegie Mellon University, Pittsburgh, PA 15213, USA}
\author[0000-0003-0691-6688]{Yang, Y.-H.}
\affiliation{Department of Physics, University of Rome ``Tor Vergata'', via della Ricerca Scientifica 1, I-00133 Rome, Italy}
\author[0009-0000-6957-8466]{Gaudin, T. M.}
\affiliation{Department of Astronomy and Astrophysics, The Pennsylvania State University, 525 Davey Lab, University Park, PA 16802, USA}
\author[0009-0004-9520-5822]{Yadav, M.}
\affiliation{Department of Physics, University of Rome ``Tor Vergata'', via della Ricerca Scientifica 1, I-00133 Rome, Italy}
\author[0009-0007-6886-4082]{Passaleva, N.}
\affiliation{Department of Physics, University of Rome ``Tor Vergata'', via della Ricerca Scientifica 1, I-00133 Rome, Italy}
\affiliation{Department of Physics, University of Rome ``Sapienza'', P.le Aldo Moro 2, I-00185 Rome, Italy}
\author[0000-0001-6849-1270]{Dichiara, S.}
\affiliation{Department of Astronomy and Astrophysics, The Pennsylvania State University, 525 Davey Lab, University Park, PA 16802, USA}

\begin{abstract}
Short gamma-ray bursts (GRBs) at redshift $z\,\gtrsim$2 remain exceptionally rare, yet they are crucial for tracing compact binary mergers in the early Universe and understanding their role in the production of r-process elements. 
GRB 061201 is an unusual and still debated event: although its optical afterglow was accurately localized, no secure coincident host galaxy was identified, and the proposed associations
with nearby galaxies all require a large separation between the GRB and its birth site. In this work, we revisit GRB\,061201 and argue that the observations are more naturally explained if the burst occurred within a
faint $F322W2$\,$\sim$\,28.4\,AB mag galaxy at $z$\,$\gtrsim$2. 
By combining constraints from the afterglow and deep near-infrared imaging from \textit{JWST}, we show that a distant origin 
($z$\,$\approx$\,3\,--\,4) provides a coherent explanation of the burst phenomenology. If confirmed, GRB 061201 would represent one of the most distant short GRBs known, extending the observed compact merger population to an epoch when the Universe was only about two billion years old. 
\end{abstract}


\section{Introduction} \label{sec:introduction}
Mergers of two neutron stars (NSs) are loud sources of gravitational waves (GWs;~\citealt{Abbott17b, Abbott2020}), progenitors of gamma-ray bursts (GRBs; \citealt{Eichler89,1991AcA....41..257P,Fryer1999,Rezzolla2011,Abbott2017a}) and the only confirmed cosmic source of heavy r-process metals \citep[e.g.][]{Symbalisty1982,Freiburghaus1999, Pian2017, Kasen2017,Domoto2022,Hotokezaka2023,Yang2024}. 
However, their poorly constrained age distribution does not allow us to understand their role in the chemical evolution of our universe~\citep{Beniamini2016,Cote2019,Kobayashi2020,Cowan2021}. 

To establish whether compact binary mergers are one of the main, if not the primary, sources of heavy r-process elements, it is necessary to estimate how common they are across cosmic time. A prompt channel of mergers, characterized by short ($\lesssim$100 Myr) lifetimes, is a prediction of stellar evolution models~\citep{Belczynski2006,Beniamini2016,Tauris2017,Beniamini2019,Andrews2019}; however, its contribution relative to the traditional channel of delayed mergers with Gyr lifetimes is not well constrained. 
Binary stellar evolution involves many uncertain physical parameters, such as multiple phases of mass transfer, mass loss, common envelopes, and tidal interactions~\citep{Dominik2012,Postnov2014,Kruckow2018}. All these factors may affect the initial orbital separation of the two NSs; thus, their lifetime as a compact binary: the tighter the orbit, the shorter the inspiral time. Population synthesis models predict a wide range of possible inspiral times depending on the details of the simulation, from $\approx$10 Myr \citep[e.g.][]{Andrews2020, Beniamini2024} to $\gtrsim$10 Gyr \citep[e.g.][]{VignaGomez2018}. These uncertainties translate into a large uncertainty in the theoretical redshift distribution of NS mergers, especially at redshifts $z>$2 where few observational constraints currently exist. 

The primary channel to pinpoint these mergers at cosmological distances are short GRBs and their afterglows. 
In the past two decades, thanks to the Neil Gehrels \textit{Swift} satellite \citep[hereafter \textit{Swift,}][]{Gehrels04} and its rapid localizations, great strides have been made in determining the distance scale and typical environment of short GRBs \citep[e.g.][]{Gehrels05,Barthelmy2005,Nicuesa2012,Fong2013,Pandey2019,Fong2022,OConnor2022,Im2024}. 
However, due to observational biases, the number of events above $z$\,$>$\,2 remains limited \citep[see, e.g.,][]{OConnor2022}. 

In this context, we revisit the case of GRB\,061201, focusing on the unsettled issue of its distance scale.
This GRB was classified as a hostless burst~\citep{Berger2010} as no coincident host galaxy was identified in deep optical and near-infrared (nIR) imaging, and 
on probabilistic grounds, there are multiple candidate host galaxies in the field.  
\citet{Stratta2007} discussed the association with a nearby galaxy (hereafter G1) at $z$\,$\sim$\,0.111, located $\approx$16\arcsec\ away from the GRB position. 
\citet{Fong2013} and, more recently, \citet{Mao2026}, presented another candidate host galaxy (hereafter G2), located $\approx$2\arcsec\ away and at a distance of $z\approx1.2$. 
However, these prior works relied on the tentative detection of an ultraviolet (UV) counterpart to rule out any galaxy  $z\gtrsim$1.7 \citep{Stratta2007}. 

Here we discuss a third possibility, that GRB\,061201 was harbored in a faint galaxy (hereafter G3) underlying the afterglow position and likely located at $z$\,$\approx$\,3-4, which would represent the most distant NS merger localized to date. Our analysis of the \textit{Swift} Ultraviolet Optical Telescope (UVOT; \citealt{Roming2005}) data shows that this interpretation is not excluded by the UV observations, which provide only weak constraints on the distance scale.

The paper is organized as follows: in Sect. 2 we summarize the existing observations, detail our re-analysis of the \textit{Swift} UVOT dataset, and present new constraints from near-infrared imaging of the field; in Sect. 3 we revisit
the host galaxy association and discuss its implications for the GRB distance scale and the rate of events. 
Unless otherwise stated, uncertainties are quoted at the $1\sigma$ confidence level for one parameter of interest \citep{Lampton1976}, and upper limits are reported at the $3\sigma$ level. We adopt a flat $\Lambda$CDM cosmology with $H_0 = 67.4~{\rm km~s^{-1}~Mpc^{-1}}$ and $\Omega_{\rm m}=0.315$ \citep{Planck2020}.

\section{Observations and Data Analysis} \label{sec:observations}

GRB~061201 is a short duration ($T_{90}=0.8\pm0.1$~s; 15-350 keV), 
hard spectrum ($E_{\rm p}=900^{+500}_{-300}$~keV) burst discovered by 
the Burst Alert Telescope (BAT; \citealt{Barthelmy2004}) aboard \textit{Swift} at  $T_0=\mathrm{2006~Dec~1~15{:}58{:}36}$~UT,  
and independently triggered by Konus-Wind \citep{Svinkin2016}. 
Its multi-wavelength follow-up campaign with \textit{Swift} and the 
ESO Very Large Telescope (VLT) was discussed in detail in \citet{Stratta2007}.
We present a re-analysis of the \textit{Swift}/UVOT images along with data acquired with the Southern Astrophysical Research (SOAR) Telescope 
and \textit{JWST}, 
which have a critical role in setting the gamma-ray burst distance scale. 
X-ray data products were retrieved from the online \textit{Swift} repository \citep{Evans2007}.

\subsection{Swift/UVOT} \label{sec:uvot}

UVOT began settled observations of the field at 
$T_0+ 86$~s and imaged the field in all six optical and ultraviolet filters. 
A detection in the broad white ($wh$) filter was reported, along with low-significance signals in the UV and optical $u$ filters. 
The marginal detection in the bluer $uvw2$ filter was used to set an upper bound of $z\lesssim1.7$ on the distance scale \citep{Stratta2007}.

We processed the images using the High Energy Astrophysics Software (HEASoft) tools version 6.33. 
In our re-analysis, we checked each snapshot for common issues that can be found in UVOT images, such as smeared stars \citep{2022Modiano} and small-scale sensitivity variations found on the detector \citep{2010Breeveld}.
No smeared snapshots were detected, but many of the UV frames are flagged as the target position lies in detector regions affected by known small-scale sensitivity variations. 
After excluding these flagged frames, no UV detection is found in any filter.

Even setting aside instrumental systematics (option \texttt{sssfile}=none), a $uvw2$ detection is difficult to reconcile with the constraints at other wavelengths.
We recover an excess in $uvw2$ at the 2.9\,$\sigma$ level
(statistical uncertainty, single trial). However, despite the longer exposure and larger effective area of other filters, we do not recover any other marginal detection, and  instead place the upper limits listed in Table~\ref{tab:photometry}.

A possible detection in $wh$ has a higher nominal significance, $\approx$4\,$\sigma$ (statistical uncertainty, single trial).
However, in the $u$ , $b$, $v$ and $wh$ images, a read-out streak from a nearby bright star crosses the target position (see the left panel of Figure \ref{fig:uvot} in Appendix \ref{appendix:uvot}), so these measurements and their significance should be treated with caution.
Similar caution applies to the $uvw1$ observations in segments 000 and 001, which show scattered light rings overlapping with the source position (Figure~\ref{fig:uvot}, right panel). 

Our re-analysis therefore supports only a weak detection in the $wh$ filter, with no convincing evidence for an associated UV counterpart.
Given the broad wavelength coverage of the $wh$ filter, the resulting constraints from the Lyman limit is only $z$\,$\lesssim$5.6. For reference, the most distant burst
detected by UVOT in the $wh$ filter is GRB\,060522 at $z$\,$\approx$5.1 \citep{GCN060522}.

\begin{table}
\centering
\caption{Photometric observations of GRB\,061201. Values are not corrected for Galactic extinction.}
\label{tab:photometry}
\begin{tabular}{lcccc}
\hline
\(T-T_0\) & Instrument & Filter & Exposure & Magnitude \\
(s) &   & & (s) & (AB) \\
\hline
135    & \textit{Swift}/UVOT & \(wh\) & 98   & $>$\,21.5 \\
389    & \textit{Swift}/UVOT & \(v\)  & 393   & $>$\,20.1 \\
5892   & \textit{Swift}/UVOT & \(u\)  & 197   & $>$\,21.2 \\
6097   & \textit{Swift}/UVOT & \(b\)  & 197   & $>$\,20.3 \\
6301   & \textit{Swift}/UVOT & \(wh\) & 197   & 21.8$\pm$0.3 \\
9894   & \textit{Swift}/UVOT & \(v\)  & 124   & $>$\,20 \\
12127  & \textit{Swift}/UVOT & \(v\)  & 590   & $>$\,21.2 \\
17770  & \textit{Swift}/UVOT & \(w1\) & 303   & $>$\,21.9 \\
41220 & \textit{Swift}/UVOT & \(w1\) & 11755 & $>$\,23.9 \\
130211& \textit{Swift}/UVOT & \(w1\) & 19972 & $>$\,24.2 \\
31018 & VLT/FORS2\tablenote{VLT/FORS2 magnitudes are taken from \citet{Stratta2007} and converted into the AB system using $R_{\rm AB}=R_{\rm Vega}+0.21$ and $I_{\rm AB}=I_{\rm Vega}+0.44$.}& \(I\) & 1080 & 22.80 $\pm$ 0.08 \\
32227 & VLT/FORS2$^{\rm a}$ & \(R\) & 540 & 23.31 $\pm$ 0.12 \\
119145 & VLT/FORS2$^{\rm a}$ & \(I\) & 1800 & $>$24 \\
41183 & SOAR/OSIRIS & \(J\) & 4260 & 22.5 $\pm$ 0.2 \\
48895  & SOAR/OSIRIS & \(K_s\) & 1440 & $>$\,22 \\
123711  & SOAR/OSIRIS & \(J\) &  5400 & $>$\,23.6 \\
\hline
\end{tabular}
\end{table}

\begin{figure*}
    \centering
    \includegraphics[width=0.98\linewidth]{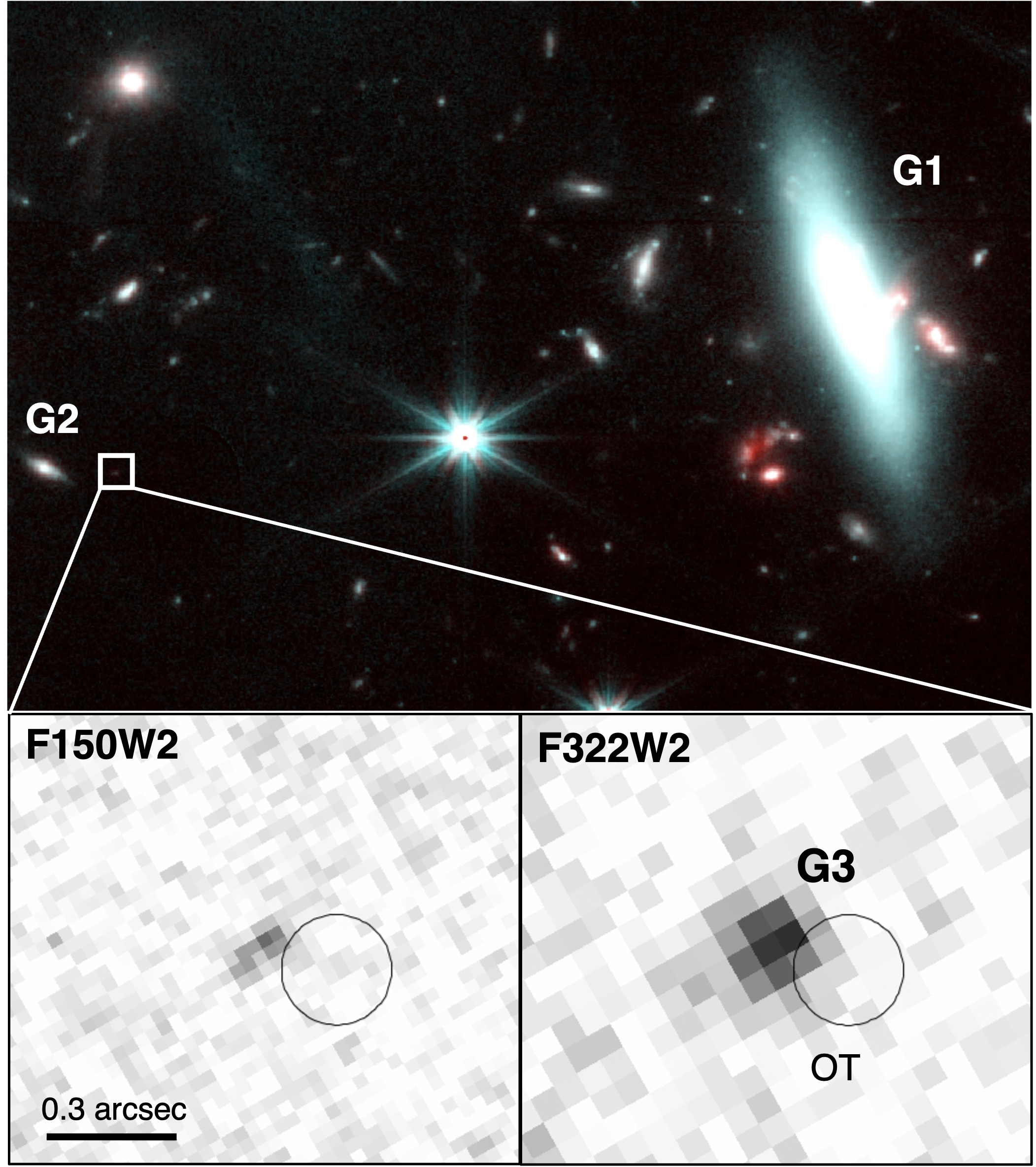}
    \caption{The field of GRB061201 observed with \textit{JWST}/NIRCam using the $F150W2$ (blue/green) and $F322W2$ (red) filters. Three galaxies are considered as possible host galaxies: G1 at $z\sim0.111$, G2 at $z\sim1.2$, and G3 at $z\gtrsim2$. The insets zoom in on the target position, which lies 0.21\arcsec $\pm$ 0.13\arcsec (68\% confidence level) from the centroid of G3. }
    \label{fig:jwst}
\end{figure*}

\subsection{Near-Infrared Imaging}
\subsubsection{SOAR}\label{sec:soar}

The target was observed on two consecutive nights with the Ohio State Infrared Imager/Spectrometer (OSIRIS), mounted on the 4.1 m Southern Astrophysical Research (SOAR) Telescope. The first epoch, obtained at $T-T_0 \approx 0.48$ d, consisted of $J$-band imaging (at an average airmass of 1.9) followed by $K_s$-band imaging (at an average airmass of 2.5).  A second deeper $J$-band sequence was obtained at $T-T_0 \approx 1.43$ d
at an average airmass of 1.8. 

Science exposures and calibration files were retrieved from the NOIRLab Astro Data Archive and reduced with a custom Python-based pipeline following standard image-processing steps, including bias subtraction, dark correction, flat-fielding, and astrometric alignment.
The field was photometrically calibrated using two bright stars from the 
VISTA Hemisphere Survey (VHS) catalog DR5 \citep{VISTADR5}
and converted into the AB system using $J_{\rm AB}=J_{\rm Vega}+0.916$ and 
$K_{s,\rm AB}=K_{s,\rm Vega}+1.839$.

As the GRB counterpart is faint, we perform aperture photometry using a circular region of radius 0.8$\times$FHWM $\approx$\,0.8\arcsec. 
Our analysis confirms the detection of a nIR counterpart  \citep{SOARGCN} with $J$\,=\,22.5\,$\pm$\,0.2 AB mag that rapidly fades between the two epochs (Table~1). 

\subsubsection{JWST/NIRCam}
\label{sec:jwst}

The field of GRB\,061201 was imaged with \textit{JWST}/NIRCam 
using the wide  $F150W2$ and $F322W2$ filters on 2023 September 27 (Figure~\ref{fig:jwst}), approximately 16.8 yr after the burst. 
Each image consisted of four dithered exposures with a total exposure time of $\approx$7.6 ks.

We retrieved the public level-3 calibrated data from Mikulski Archive for Space Telescopes (MAST). 
To place the JWST images on the same astrometric frame as the afterglow, we registered them to the VLT $I$ band image \citep{Stratta2007} using 
Source Extractor \citep{Bertin1996} for source detection and SCAMP for alignment~\citep{Bertin2006}. The astrometric solution was derived from 24 common sources, which yield a rms residual offset of 0.098\arcsec.
At the GRB position, the NIRCam imaging reveals a faint, red source, hereafter referred to as G3 (Figure~\ref{fig:jwst}).  We measure an angular separation of \(0.21\arcsec \pm 0.13\arcsec\) between G3 and the GRB position, including the statistical uncertainty on both the position of the GRB afterglow, the faint JWST source G3, and the astrometric tie uncertainty of the two images.

We performed aperture photometry adopting a 
circular aperture with radius 0.2\arcsec~centered on G3 and estimating the local background from a nearby source-free annulus. Aperture corrections were applied using the NIRCam encircled-energy curves, and fluxes were converted to AB magnitudes using the photometric calibration provided in the image headers.
The derived values are $F150W2=$29.8$\pm$0.3 and $F322W2=$28.42$\pm$0.15 AB mag.

With only two photometric points, the red color, 
\(F150W2-F322W2 \sim 1.4\), does not provide a unique constraint on either the photometric redshift or the dust content of the galaxy. Several short GRB host galaxies show prominent Balmer/4000~\AA\ breaks \citep{Leibler2010,OConnor2022}. If the observed color is produced by such a break falling within the $F150W2$ bandpass, then the redshift would be bracketed to approximately \(2 \lesssim z \lesssim 5\). If emission lines
significantly contribute to the flux in \(F322W2\), then G3 may be located at even higher redshifts. For instance, the Balmer H$\alpha$ line enters the \(F322W2\) bandpass at \(z \sim 2.7\)--\(2.8\),  and the O[III] doublet at \(z \gtrsim 4\). 
Therefore, although the distance scale of G3 cannot be constrained with
this dataset, there exists a range of plausible solutions that can explain
its red color while remaining consistent with the GRB afterglow limits.

\section{Results}\label{sec:res}
\subsection{Host galaxy association }

GRB 061201 has traditionally been associated with a galaxy (G1 in Figure~\ref{fig:jwst}) at $z \approx$ 0.111 \citep{Stratta2007}, which is offset by $\approx$\,16.2\arcsec ($\approx$32.5 kpc in projection) from the GRB localization \citep{Fong2013}. 
This association was supported by a relatively low probability of chance coincidence, \textbf{$P_{cc} \approx 11\%$} in $r$ band (see Table~\ref{tab:pcc_candidates}), and by the UV constraints on the afterglow redshift suggesting a low-redshift origin,  $z\,\lesssim$1.7~\citep{Stratta2007}.


Our re-analysis of the UVOT data significantly relaxes this constraint. We find that the positive fluctuations observed in the optical and UV filters are likely induced by instrumental effects, including read-out streaks and scattered-light rings (Appendix \ref{appendix:uvot}, Figure \ref{fig:uvot}). The detection in the broad $wh$ filter would then sets $z\,\lesssim$\,5.6. 

A more stringent limit is derived from the VLT optical spectroscopy~\citep{Stratta2007}. By inspecting the 2D spectrum, a weak trace is clearly visible at the GRB position down to $\approx$5100\,\AA\ (Appendix \ref{appendix:vlt}). Below this wavelength, the background noise becomes too high to securely identify a source. The resulting upper bound from the Lyman limit is $z\,\lesssim$\,4.6. 

In light of these new findings, we consider three potential GRB/galaxy associations: G1 at $z\approx$0.111, G2 at $z\approx$1.2,  and G3 at $z\gtrsim$2. 
Based on the probability of chance coincidence $P_{cc}$ \citep{Bloom2002}, 
no clear candidate emerges (Appendix~\ref{appendix:pcc}, Table~\ref{tab:pcc_candidates}):
$P_{cc}$ ranges from 5\% to 22\%, and the most likely association also depends on the selected filter. We therefore examine the GRB properties and assess how well they are reproduced under each proposed association.

Several lines of evidence disfavor the association with G1. 
First, the faint and rapidly fading optical emission is not consistent with the behavior of an AT2017gfo-like kilonova at $z\,\sim\,0.1$ (cf. Figure 4 of \citealt{Troja2023}). 
This conclusion is strengthened by the nIR constraints: at $z \sim 0.1$, an AT2017gfo-like kilonova would peak at $J \approx 23$ AB mag, above the limits derived in Table~1.

A further inconsistency arises from the large projected offset of $\approx$32 kpc between G1 and the GRB position. At such a distance from the putative host, the burst would be expected to occur in a low-density environment and produce a faint X-ray afterglow~\citep{Kumar2000,OConnor2020}. Instead, the observed X-ray flux is commensurate with the energy of the explosion. 

We derive an X-ray to prompt gamma-ray ratio of 
log\,(\(F_{\rm X,11hr}/S_{\gamma}) \approx -6.1\), 
where \(S_{\gamma}\) is the prompt gamma-ray fluence in the 15-150 keV band and \(F_{\rm X,11hr}\) is the X-ray flux in the 0.3-10 keV band evaluated at 11 hr. This quantity provides a useful proxy of the circumburst density \citep{Nysewander2009,OConnor2022}, and its value favors a modest physical offset ($\lesssim$10 kpc, cf. Figure 1 of \citealt{Yang2024}). For comparison, two large-offset bursts, GRB\,211211A and GRB\,230307A, have values of \(\approx-7.9\)~\citep{Troja2022} and \(\approx-8.3\)~\citep{Yang2024}, respectively.

\begin{figure*}
    \centering
    \includegraphics[width=0.45\linewidth]{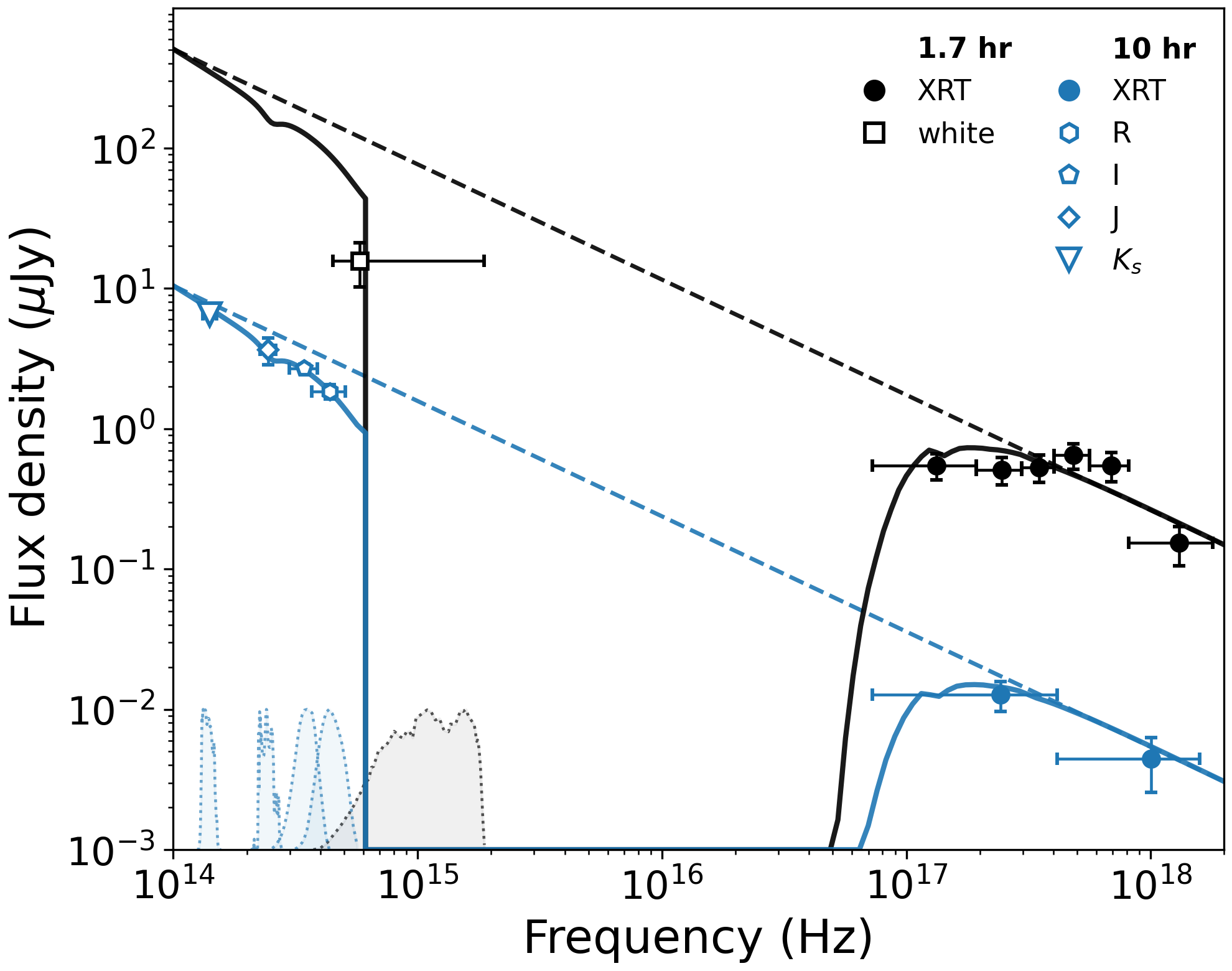} \hspace{0.2cm}
    \includegraphics[width=0.45\linewidth]{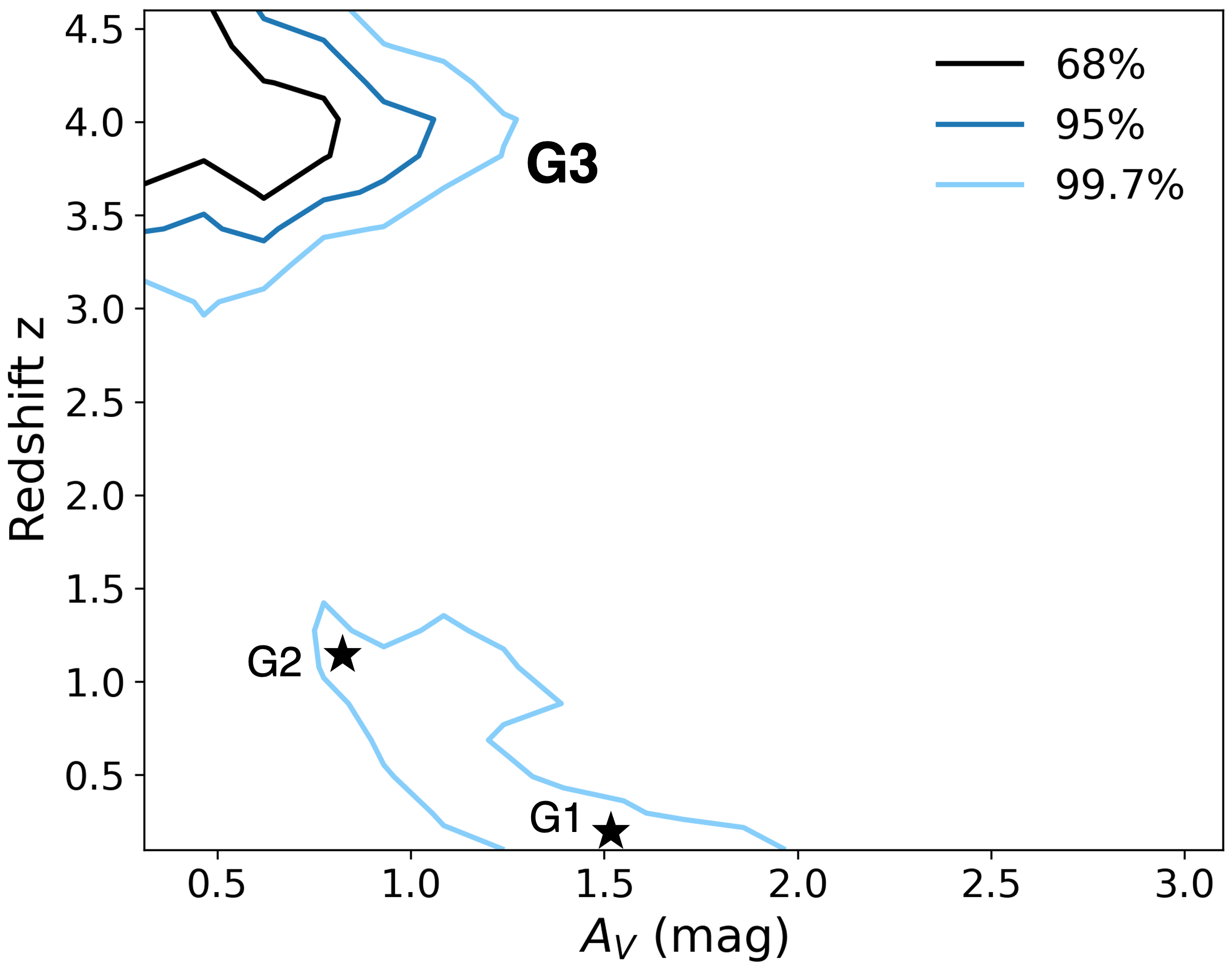}
    \caption{\textit{Left:} Broadband afterglow SED at $T_0+1.7$\,hr and $T_0+10$\,hr. 
    X-ray data (filled circles) and optical/nIR photometric points (open symbols) are best fit by a 
    power-law model with index $\beta\approx$0.8, shown  by the dashed line (without absorption/extinction) and by the solid line (including absorption and extinction effects). The dotted shaded curves show the filter response functions. \textit{Right:} Confidence contours for intrinsic extinction $A_V$ and redshift $z$. Contours show the 68\%, 95\%, and 99.7\% confidence regions. The favorite model corresponds to a high-redshift origin, consistent with an association with G3. The stars indicate the best fit values for G1 and G2.}
    \label{fig:sed}
\end{figure*}

\citet{Mao2026} argued in favor of an association with G2 for which 
they derived a photometric redshift $z$\,$\approx$\,1.2. 
This redshift lies close to the peak of the NS merger rate distribution \citep{Beniamini2019}, where a substantial fraction of short GRBs are expected to occur. However, this association would still imply a large projected physical offset ($\approx$17 kpc) which, as for G1, is not supported by the X-ray afterglow  brightness. 

Instead, we propose G3 as the most viable candidate.
Based on chance coincidence probability $P_{cc}$ (see Table~\ref{tab:pcc_candidates}), G3 is broadly equivalent to G1 and G2 in a statistical sense. 
An association with G3 remains consistent with the revised UV/optical constraints, does not imply a large separation or a high kick velocity, and alleviates the tension with the X-ray afterglow brightness. 
    Although G3 is extremely faint in apparent magnitude, its brightness would fall within the typical distribution of short GRB host galaxies if G3 lies at $z$\,$\gtrsim$\,3 \citep{Fong2013,Dichiara2026}, a redshift range also consistent with its red color (Sect.~\ref{sec:jwst}).
Additionally, we argue that its possible higher redshift would better account for the observed afterglow spectral energy distribution (SED). 

\subsection{A high-redshift origin for GRB\,061201}

We fit the broadband afterglow SEDs in count space using an absorbed power-law model, including two components for
gas absorption and two components for dust extinction (\texttt{TBabs*zTBabs*redden*zdust*powerlaw} within XSPEC, \citealt{Arnaud1996}) to model the Galactic and host galaxy contributions.
The redshift is left free to vary between $z_{\min}$=0.1, set by G1, and  $z_{\max}$=4.6, constrained by the VLT spectrum (see Appendix~\ref{appendix:vlt}). 

In Figure~\ref{fig:sed}, we show the afterglow SEDs at two different epochs, \(T_0+1.7\) hr and \(T_0+10\) hr, along with the best fit models. 
At \(T_0+10\) hr, the spectral slope is well constrained by the X-ray, optical, and nIR detections.  At earlier times, the X-ray emission is an order of magnitude brighter, yet the simultaneous UVOT observations  yield a weak detection, \(wh \approx 21.8\) AB mag, which falls well below the extrapolation of the best-fit power-law (dashed line).
\citet{Stratta2007} interpreted the observed SED in terms of spectral evolution, with the optical-to-X-ray slope changing from \(\beta \approx 0.5\) at \(T_0+1.7\) hr to \(\beta \approx 0.8\) \(T_0+10\) hr. However, such a spectral change is not accompanied by any obvious change in the temporal behavior. Within the standard synchrotron afterglow model, it would also require a rapid passage of the cooling frequency with \(\nu_c \propto t^{-4}\), ranging
from \(\nu_c \gtrsim 1\) keV at 1.7 hr to \(\nu_c \lesssim 1\) eV at 10 hr. This is extremely steep compared with standard afterglow expectations.

Under the assumption of no spectral evolution between 1.7 hr and 10 hr, the faint $wh$ flux can be explained either by dust extinction or a higher redshift (Figure~\ref{fig:sed}, right panel).  
An obscured line of sight ($A_V$\,$\gtrsim$1) at redshift $z\lesssim$1.5 is consistent with the afterglow spectrum but is at odds with the large galacto-centric offset implied by an association with either G1 or G2. In both these cases, the burst position lies outside the galaxy's light, where substantial dust extinction is not expected. 

We argue instead that the faint optical/UV flux is more naturally explained by Lyman absorption, as expected if GRB\,061201 occurred at higher redshift. For \(z \sim 3\), absorption by Ly\(\alpha\) systems begins to suppress the spectrum blueward of \(\lambda \sim 4900\)\,\AA, and the flux drops below the redshifted Lyman limit at \(\lambda \sim 3700\)\,\AA,  strongly affecting the UVOT bands while leaving the red optical and nIR afterglow largely unaffected. In this scenario, the faint optical flux at early times is not evidence of intrinsic spectral evolution, but rather reflects absorption blueward of the redshifted Lyman limit.
Our best fit model, shown in Figure~\ref{fig:sed} (left panel), is described by a redshift $z$\,=\,4.3$^{+0.3}_{-0.5}$, a photon index \(\Gamma = 1.84 \pm 0.08\), an intrinsic hydrogen column density
\(N_{\rm H,z} = (1.8^{+1.6}_{-1.3})\times10^{22}\,{\rm cm^{-2}}\), and 
reddening \(E(B-V)=0.06^{+0.04}_{-0.05}\), assuming a Milky Way (MW) extinction law \citep{Pei1992}. The Galactic foreground values were kept fixed at \(E(B-V)=0.065\) and \(N_{\rm H}=6.7\times10^{20}\,{\rm cm^{-2}}\) \citep{SF2011,Willingale2013}. 
The redshift is constrained to \(z\gtrsim3.3\) at the \(2\,\sigma\) confidence level for a MW extinction law (Figure~\ref{fig:sed}, right panel), and \(z\gtrsim2.4\) at the \(2\,\sigma\) confidence level for a Small Magellanic Cloud (SMC) extinction law (see Appendix~\ref{appendix:smc}, Figure~\ref{fig:sedSMC}).

Using a scale of 8\arcsec/kpc, appropriate for this range of redshifts, 
the observed offset corresponds to a physical projected distance of 
$1.7\pm1.0$\,kpc. 
Assuming $z\,\approx\,3.5$, the rest-frame peak energy and isotropic-equivalent gamma-ray energy would shift to $E_{p,z}$=4.1$^{+2.3}_{-1.4}$ MeV and 
$E_{\gamma,\rm iso}$\,$\approx$10$^{53}$\,erg, respectively. These values lie at the high end of the short GRBs distribution but remain within the observed range and consistent with the location of short 
GRBs in the Amati diagram \citep{Amati2002}.

\subsection{Implications for the rate of events}

Over the past 20 years, \textit{Swift} has delivered rapid localizations for more than 130 short GRBs, enabling the first census of their properties, environment and distance scales. 
However, the tails of the observed distribution, the very nearby events ($z\lesssim$0.2) and the most distant ones ($z\gtrsim$2), remain poorly populated. 
If associated with G1 at \(z\sim0.111\), GRB\,061201 would belong to the small sample of nearby short GRBs 
together with  GRB150101B, GRB160821B, GRB080905A and GRB130822A\footnote{We caution that the redshift of GRB130822A is based on the association with a nearby galaxy $\approx$\,20\arcsec~away, with $P_{cc}\approx$8\% \citep{OConnor2022}}), and would contribute to estimates of the local merger rate. Removing it from this low-redshift bin reduces the sample (hence the rate density) by 20\%, helping to ease the mild tension with GW-based rates \citep{2026arXiv260405046K}.
As a simple estimate, we compute the local rate of nearby short GRBs using a \(V_{\max}\) method \citep[e.g.][]{Jin2018,Troja2022}:
\begin{equation}
\mathcal{R}_{z<0.2} \approx 
\frac{4\pi N}{\Delta t\,\Omega\,V_{\max}\,f_b} \approx 0.6\,f_b^{-1}\,{\rm Gpc}^{-3}\,{\rm yr}^{-1},
\end{equation}
where \(N=4\) is the number of events in the sample, \(\Delta t \approx 16\) yr is the effective observing time, corresponding to a \(\sim80\%\) duty cycle over the mission lifetime, \(\Omega \sim 2.2\) sr is the \textit{Swift} field of view, \(V_{\max}\approx2.3\,{\rm Gpc}^3\) is the comoving volume enclosed by \(z=0.2\), and \(f_b\) is the beaming fraction. 
We do not include hybrid long GRBs \citep[e.g.][]{Gehrels06, Troja2022, Yang2024}  in this estimate, as their beaming factor and progenitor systems may be different from standard short GRBs \citep{Troja2008,Yang2022}. However, we note that their inclusion would increase the rate by a factor  $\lesssim$\,2.

\begin{figure}[!t]
    \centering
    \includegraphics[width=\linewidth]{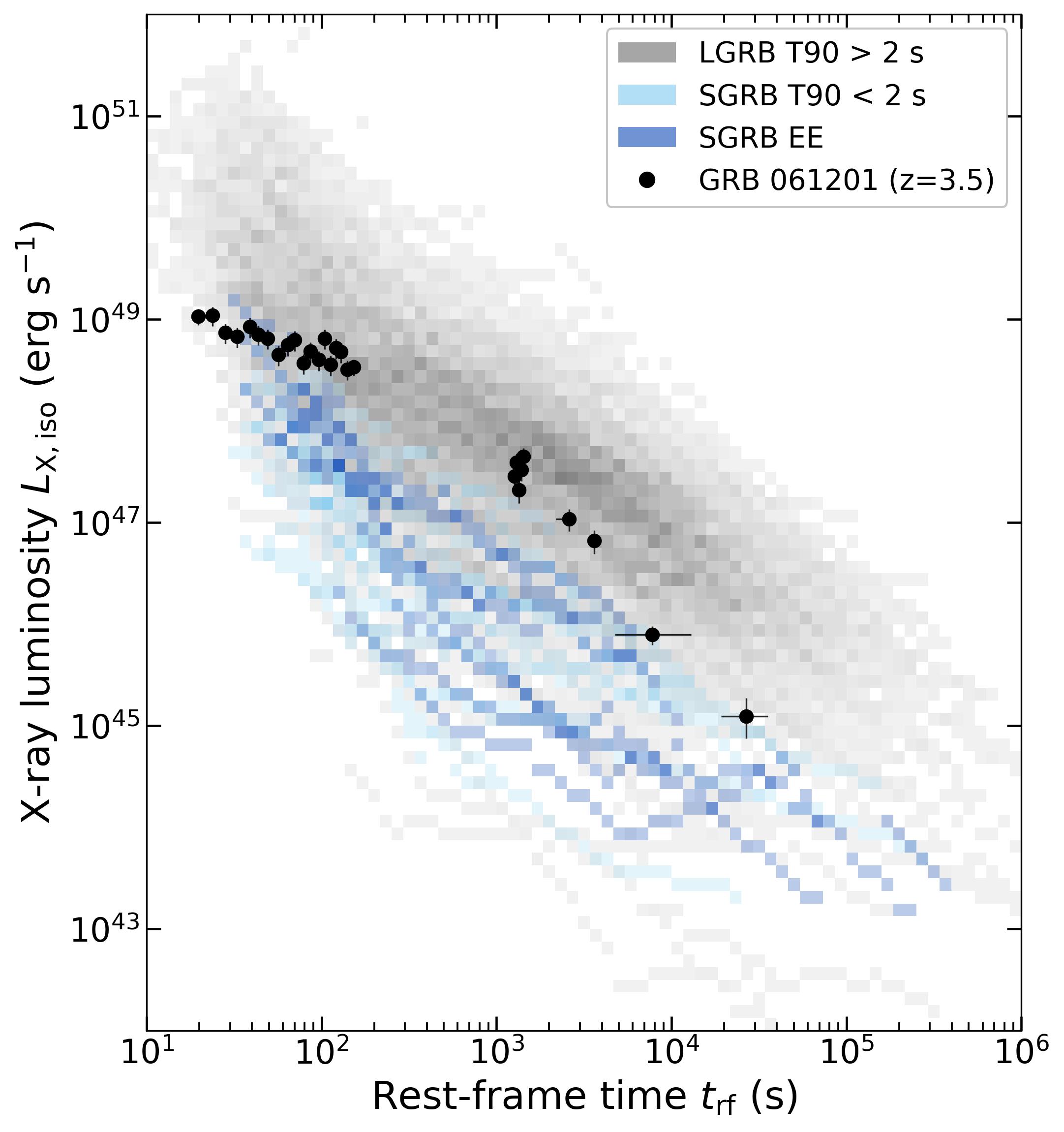}
    \caption{X-ray luminosity light curve for GRB 061201 placed at $z=3.5$ (black symbols) compared to the density maps for long GRBs (gray), short GRBs (light blue) and short GRBs with extended emission (blue).}
    \label{fig:lc}
\end{figure}

To reconcile Eq.~1 with the binary NS merger rate inferred from GWs
(\(22-250\,{\rm Gpc}^{-3}\,{\rm yr}^{-1}\); \citealt{2025arXiv250818083T}), 
a beaming correction of \(f_b^{-1}\approx 40-400\), corresponding to jet opening angles of \(\theta_j\sim4^\circ-12^\circ\), would be sufficient, provided that all binary NS mergers launch successful short GRB jets. 
This requirement is softened by the statistical uncertainty associated with the small number of events: for \(N=4\), the 90\% credible range corresponds to a rate of \((0.2-1.5)\,f_b^{-1}\,{\rm Gpc}^{-3}\,{\rm yr}^{-1}\). After allowing for this uncertainty, the comparison suggests that a sizable fraction of binary NS mergers ($\gtrsim$30\%) must produce successful relativistic jets.

Shifting GRB\,061201 to higher redshifts also eases the requirement on its beaming angle. Its X-ray afterglow light curve is well described by a shallow power-law decay with $\alpha$\,$\sim$0.54, followed by 
a steeper decay after $\sim$\,0.7 hr (observer's frame). This behavior has often been interpreted as a jet-break, implying a very narrow jet $\theta_j \approx$1$^\circ$ \citep{Stratta2007}. 
As a result, this one event would drive the beaming-corrected rate of events to much higher values \citep[e.g.][]{Jin2018}. 
However, if placed at higher redshift, its early X-ray emission resembles, in luminosity and timescale, the tail of temporally extended  emission observed in some short GRBs \citep{Norris2006} rather than standard forward shock radiation. 
In Figure~\ref{fig:lc}, we compare the X-ray luminosity light curve of GRB\,061201, assuming a $z=3.5$ with the population of X-ray afterglows from 
long GRBs (gray), short GRBs (pale blue) and short GRB with extended emission (blue). At this high redshift, GRB\,061201 falls in the same luminosity-time region occupied by short GRBs with extended emission, suggesting that its early X-ray afterglow and subsequent flux decay do not require an exceptionally narrow jet, but can be interpreted as the result of long-lived activity of the central engine.

A higher redshift for GRB\,061201 not only affects estimates of the local event rate, but has broader implications for the inferred redshift distribution of compact binary mergers and their typical delay times. 
Experience has shown that short GRBs at $z$\,$\gtrsim$2 are difficult to detect \citep{Moss2022}, 
accurately localize \citep[e.g.][]{Sakamoto2013}, 
and robustly classify as merger-driven bursts \citep[e.g.][]{Levesque2010, Dichiara2021}. 
Additionally, since most redshifts come from the putative host galaxies, one has to take into account that measuring the distance scale of these faint galaxies is observationally challenging. 
For example, \citealt{Nugent2022} proposed five more events (GRB\,051210, GRB\,160408, GRB\,170127B, GRB\,180727A, GRB\,191031D) at $z\gtrsim2$ based on the SEDs of their putative host galaxies. However, as in the case of GRB\,061201, these redshifts are subject to significant uncertainties in their host association and photometric estimates (cf.  \citealt{OConnor2022}, \citealt{Im2024}), requiring further investigation to confirm them.
As a consequence of these challenges, even a single additional event added to the high-$z$ sample carries significant weight and implies that the intrinsic merger rate at early epochs cannot be negligible. 

\begin{figure}
    \centering
    \includegraphics[width=\linewidth]{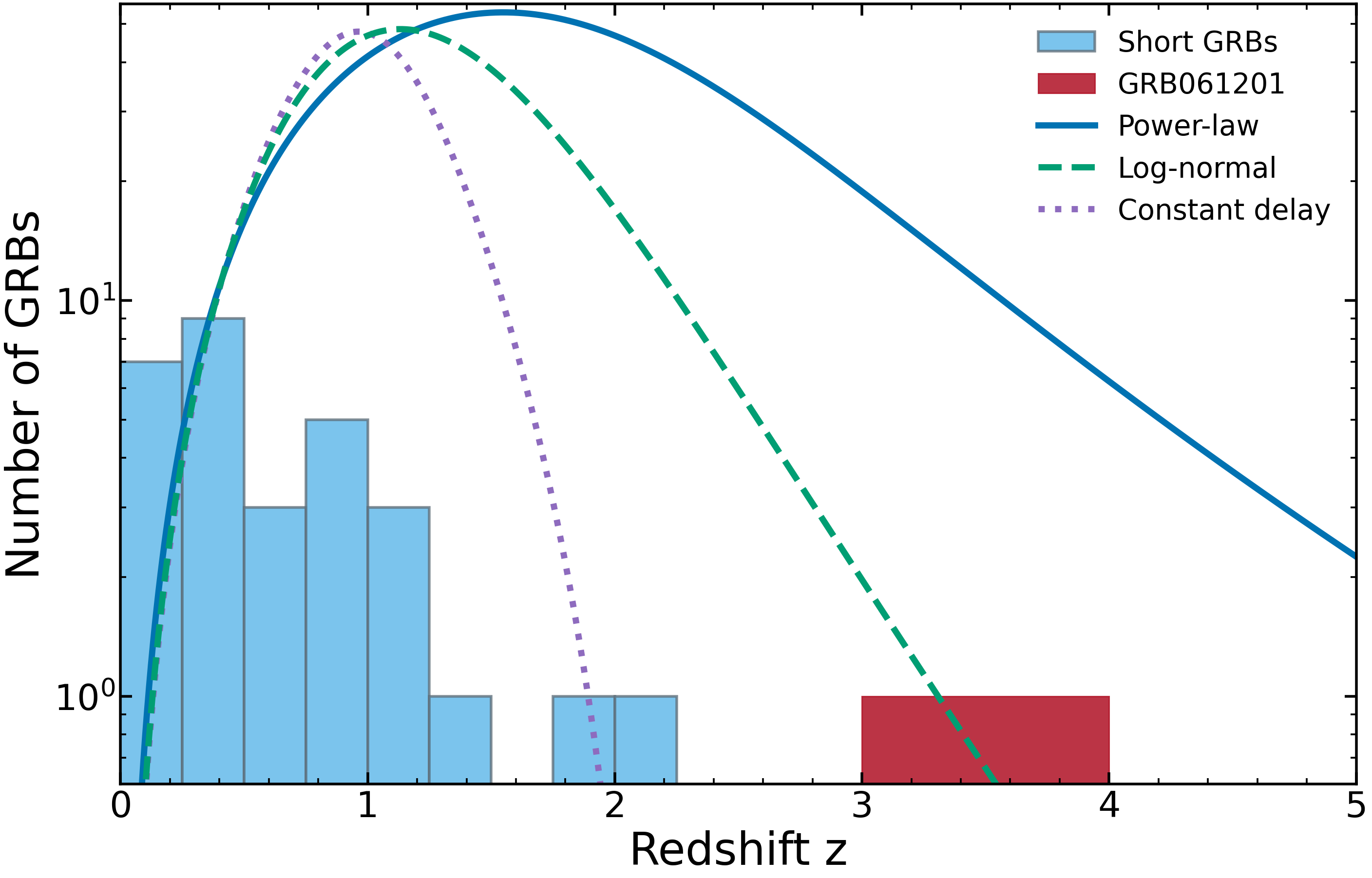}
    \caption{Observed redshift distribution of short GRBs compared with the predictions of different delay time distributions. The red bin shows the approximate location of GRB\,061201. }
    \label{fig:dtd}
\end{figure}

In Figure~\ref{fig:dtd}, we report the redshift distribution of
short GRBs, selected using the criteria $T_{90}\lesssim$\,1~s and $P_{cc}<0.2$ to minimize contamination from collapsar-driven bursts \citep{Bromberg2012} and spurious host galaxy associations. 
This is compared with different models of delay time  distributions with respect to the star-formation history \citep{MadauFragos2017}, 
such as a constant delay time $\tau\approx$\,3 Gyr (dotted line; \citealt{Wanderman2015}), a log-normal distribution
centered at $\tau\approx$\,2 Gyr  with $\sigma \approx$\,0.5 (dashed line), 
and a power-law distribution with slope $-1$ \citep{Piran1992}
and a minimum delay time of 20 Myr (solid line). 
These models are roughly equivalent at $z\lesssim$1 but their predictions differ drastically  at $z\gtrsim$2. 
Observational evidence for short GRBs at these redshifts can therefore provide strong leverage on the delay time distribution of compact binary mergers. In particular, the possible association of
GRB~061201 with a high redshift host at $z\gtrsim2$ (and likely $z$\,$\approx$\,3\,--\,4 as suggested by the afterglow modeling) 
would be difficult to reconcile with distributions dominated by long delays, while it is more naturally accommodated by a broader power-law distribution.

\newpage
\section{Conclusion}

We have revisited the distance scale and host galaxy association of
GRB~061201. Our re-analysis of the
\textit{Swift}/UVOT data shows that the source position is affected by
instrumental systematics and no secure UV counterpart is recovered.
The remaining optical limits allow for a substantially higher redshift. 

Deep \textit{JWST}/NIRCam imaging reveals a faint red source, G3,
nearly coincident with the afterglow position. 
We derive a modest chance coincidence probability,  
and present the observational evidence supporting a physical link between GRB\,061201 and G3. 
Although the available
two-band photometry does not yield a unique photometric redshift for G3, its
color and brightness are consistent with a galaxy at $z\gtrsim2$.
A high redshift  of $z$\,$\approx$\,3\,--\,4 is also supported by the 
afterglow SED. 
A secure redshift for G3, either from
multi-color imaging with narrower bandpass filters, or, ideally, from spectroscopy, is needed to confirm this scenario.
If established, GRB~061201 would be among the most distant short GRBs
known and would provide an important observational anchor for the
evolution of compact-object mergers and their contribution to
$r$-process enrichment in the early Universe.

\section*{Acknowledgments}
This work is supported by the European Research Council through the Consolidator grant BHianca (grant agreement ID~101002761).
BO gratefully acknowledges support from the McWilliams Fellowship at Carnegie Mellon University. SD acknowledges support from the National Aeronautics and Space Administration under Award No. 80NSSC22K1516.

This research made use of data and software provided by the High Energy Astrophysics Science Archive Research Center (HEASARC), a service of the Astrophysics Science Division at NASA/GSFC; SAOImageDS9, developed by the Smithsonian Astrophysical Observatory; and data supplied by the UK Swift Science Data Centre at the University of Leicester.

This work is based in part on observations made with the NASA/ESA/CSA James Webb Space Telescope, and associated with program \#3134.
The data (doi: \hyperlink{https://archive.stsci.edu/doi/resolve/resolve.html?doi=10.17909/p5dz-zf74}
{10.17909/p5dz-zf74}) were obtained from the Mikulski Archive for Space Telescopes at the Space Telescope Science Institute, which is operated by the Association of Universities for Research in Astronomy, Inc., under NASA contract NAS 5-03127 for JWST. 

\newpage

\bibliography{main}{}

@ARTICLE{Svinkin2016,
       author = {{Svinkin}, D.~S. and {Frederiks}, D.~D. and {Aptekar}, R.~L. and {Golenetskii}, S.~V. and {Pal'shin}, V.~D. and {Oleynik}, Ph. P. and {Tsvetkova}, A.~E. and {Ulanov}, M.~V. and {Cline}, T.~L. and {Hurley}, K.},
        title = "{The Second Konus-Wind Catalog of Short Gamma-Ray Bursts}",
      journal = {\apjs},
         year = 2016,
        month = may,
       volume = {224},
       number = {1},
          eid = {10},
        pages = {10},
          doi = {10.3847/0067-0049/224/1/10},
archivePrefix = {arXiv},
       eprint = {1603.06832},
 primaryClass = {astro-ph.HE},
       adsurl = {https://ui.adsabs.harvard.edu/abs/2016ApJS..224...10S}
}

@ARTICLE{Bloom2002,
       author = {{Bloom}, J.~S. and {Kulkarni}, S.~R. and {Djorgovski}, S.~G.},
        title = "{The Observed Offset Distribution of Gamma-Ray Bursts from Their Host Galaxies: A Robust Clue to the Nature of the Progenitors}",
      journal = {\aj},
         year = 2002,
        month = mar,
       volume = {123},
       number = {3},
        pages = {1111-1148},
          doi = {10.1086/338893},
archivePrefix = {arXiv},
       eprint = {astro-ph/0010176},
 primaryClass = {astro-ph},
       adsurl = {https://ui.adsabs.harvard.edu/abs/2002AJ....123.1111B}
}

@ARTICLE{Stratta2007,
       author = {{Stratta}, G. and {D'Avanzo}, P. and {Piranomonte}, S. and {Cutini}, S. and {Preger}, B. and {Perri}, M. and {Conciatore}, M.~L. and {Covino}, S. and {Stella}, L. and {Guetta}, D. and {Marshall}, F.~E. and {Holland}, S.~T. and {Stamatikos}, M. and {Guidorzi}, C. and {Mangano}, V. and {Antonelli}, L.~A. and {Burrows}, D. and {Campana}, S. and {Capalbi}, M. and {Chincarini}, G. and {Cusumano}, G. and {D'Elia}, V. and {Evans}, P.~A. and {Fiore}, F. and {Fugazza}, D. and {Giommi}, P. and {Osborne}, J.~P. and {La Parola}, V. and {Mineo}, T. and {Moretti}, A. and {Page}, K.~L. and {Romano}, P. and {Tagliaferri}, G.},
        title = "{A study of the prompt and afterglow emission of the short GRB 061201}",
      journal = {\aap},
         year = 2007,
        month = nov,
       volume = {474},
       number = {3},
        pages = {827-835},
          doi = {10.1051/0004-6361:20078006},
       adsurl = {https://ui.adsabs.harvard.edu/abs/2007A&A...474..827S}
}

@ARTICLE{Evans2007,
       author = {{Evans}, P.~A. and {Beardmore}, A.~P. and {Page}, K.~L. and {Tyler}, L.~G. and {Osborne}, J.~P. and {Goad}, M.~R. and {O'Brien}, P.~T. and {Vetere}, L. and {Racusin}, J. and {Morris}, D. and {Burrows}, D.~N. and {Capalbi}, M. and {Perri}, M. and {Gehrels}, N. and {Romano}, P.},
        title = "{An online repository of Swift/XRT light curves of {\ensuremath{\gamma}}-ray bursts}",
      journal = {\aap},
         year = 2007,
        month = jul,
       volume = {469},
       number = {1},
        pages = {379-385},
          doi = {10.1051/0004-6361:20077530},
archivePrefix = {arXiv},
       eprint = {0704.0128},
 primaryClass = {astro-ph},
       adsurl = {https://ui.adsabs.harvard.edu/abs/2007A&A...469..379E}
}

@ARTICLE{Bertin1996,
       author = {{Bertin}, E. and {Arnouts}, S.},
        title = "{SExtractor: Software for source extraction.}",
      journal = {\aaps},
         year = 1996,
        month = jun,
       volume = {117},
        pages = {393-404},
          doi = {10.1051/aas:1996164},
       adsurl = {https://ui.adsabs.harvard.edu/abs/1996A&AS..117..393B}
}

@INPROCEEDINGS{Bertin2006,
       author = {{Bertin}, E.},
        title = "{Automatic Astrometric and Photometric Calibration with SCAMP}",
    booktitle = {Astronomical Data Analysis Software and Systems XV},
         year = 2006,
       editor = {{Gabriel}, C. and {Arviset}, C. and {Ponz}, D. and {Enrique}, S.},
       series = {Astronomical Society of the Pacific Conference Series},
       volume = {351},
        month = jul,
        pages = {112},
       adsurl = {https://ui.adsabs.harvard.edu/abs/2006ASPC..351..112B}
}

@ARTICLE{GCN060522,
       author = {{Cenko}, S.~B. and {Berger}, E. and {Djorgovski}, S.~G. and {Mahabal}, A.~A. and {Fox}, D.~B.},
        title = "{GRB060522: z = 5.11 Keck-LRIS redshift.}",
      journal = {GRB Coordinates Network},
         year = 2006,
        month = jan,
       volume = {5155},
        pages = {1},
       adsurl = {https://ui.adsabs.harvard.edu/abs/2006GCN..5155....1C}
}

@ARTICLE{Hogg1997,
       author = {{Hogg}, David W. and {Pahre}, Michael A. and {McCarthy}, James K. and {Cohen}, Judith G. and {Blandford}, Roger and {Smail}, Ian and {Soifer}, B.~T.},
        title = "{Counts and colours of faint galaxies in the U and R bands}",
      journal = {\mnras},
         year = 1997,
        month = jun,
       volume = {288},
       number = {2},
        pages = {404-410},
          doi = {10.1093/mnras/288.2.404},
archivePrefix = {arXiv},
       eprint = {astro-ph/9702241},
 primaryClass = {astro-ph},
       adsurl = {https://ui.adsabs.harvard.edu/abs/1997MNRAS.288..404H}
}

@ARTICLE{Lampton1976,
       author = {{Lampton}, M. and {Margon}, B. and {Bowyer}, S.},
        title = "{Parameter estimation in X-ray astronomy.}",
      journal = {\apj},
         year = 1976,
        month = aug,
       volume = {208},
        pages = {177-190},
          doi = {10.1086/154592},
       adsurl = {https://ui.adsabs.harvard.edu/abs/1976ApJ...208..177L}
}

@ARTICLE{Planck2020,
       author = {{Planck Collaboration} and {Aghanim}, N. and {Akrami}, Y. and {Ashdown}, M. and {Aumont}, J. and {Baccigalupi}, C. and {Ballardini}, M. and {Banday}, A.~J. and {Barreiro}, R.~B. and {Bartolo}, N. and {Basak}, S. and {Battye}, R. and {Benabed}, K. and {Bernard}, J.-P. and {Bersanelli}, M. and {Bielewicz}, P. and {Bock}, J.~J. and {Bond}, J.~R. and {Borrill}, J. and {Bouchet}, F.~R. and {Boulanger}, F. and {Bucher}, M. and {Burigana}, C. and {Butler}, R.~C. and {Calabrese}, E. and {Cardoso}, J.-F. and {Carron}, J. and {Challinor}, A. and {Chiang}, H.~C. and {Chluba}, J. and {Colombo}, L.~P.~L. and {Combet}, C. and {Contreras}, D. and {Crill}, B.~P. and {Cuttaia}, F. and {de Bernardis}, P. and {de Zotti}, G. and {Delabrouille}, J. and {Delouis}, J.-M. and {Di Valentino}, E. and {Diego}, J.~M. and {Dor{\'e}}, O. and {Douspis}, M. and {Ducout}, A. and {Dupac}, X. and {Dusini}, S. and {Efstathiou}, G. and {Elsner}, F. and {En{\ss}lin}, T.~A. and {Eriksen}, H.~K. and {Fantaye}, Y. and {Farhang}, M. and {Fergusson}, J. and {Fernandez-Cobos}, R. and {Finelli}, F. and {Forastieri}, F. and {Frailis}, M. and {Fraisse}, A.~A. and {Franceschi}, E. and {Frolov}, A. and {Galeotta}, S. and {Galli}, S. and {Ganga}, K. and {G{\'e}nova-Santos}, R.~T. and {Gerbino}, M. and {Ghosh}, T. and {Gonz{\'a}lez-Nuevo}, J. and {G{\'o}rski}, K.~M. and {Gratton}, S. and {Gruppuso}, A. and {Gudmundsson}, J.~E. and {Hamann}, J. and {Handley}, W. and {Hansen}, F.~K. and {Herranz}, D. and {Hildebrandt}, S.~R. and {Hivon}, E. and {Huang}, Z. and {Jaffe}, A.~H. and {Jones}, W.~C. and {Karakci}, A. and {Keih{\"a}nen}, E. and {Keskitalo}, R. and {Kiiveri}, K. and {Kim}, J. and {Kisner}, T.~S. and {Knox}, L. and {Krachmalnicoff}, N. and {Kunz}, M. and {Kurki-Suonio}, H. and {Lagache}, G. and {Lamarre}, J.-M. and {Lasenby}, A. and {Lattanzi}, M. and {Lawrence}, C.~R. and {Le Jeune}, M. and {Lemos}, P. and {Lesgourgues}, J. and {Levrier}, F. and {Lewis}, A. and {Liguori}, M. and {Lilje}, P.~B. and {Lilley}, M. and {Lindholm}, V. and {L{\'o}pez-Caniego}, M. and {Lubin}, P.~M. and {Ma}, Y.-Z. and {Mac{\'\i}as-P{\'e}rez}, J.~F. and {Maggio}, G. and {Maino}, D. and {Mandolesi}, N. and {Mangilli}, A. and {Marcos-Caballero}, A. and {Maris}, M. and {Martin}, P.~G. and {Martinelli}, M. and {Mart{\'\i}nez-Gonz{\'a}lez}, E. and {Matarrese}, S. and {Mauri}, N. and {McEwen}, J.~D. and {Meinhold}, P.~R. and {Melchiorri}, A. and {Mennella}, A. and {Migliaccio}, M. and {Millea}, M. and {Mitra}, S. and {Miville-Desch{\^e}nes}, M.-A. and {Molinari}, D. and {Montier}, L. and {Morgante}, G. and {Moss}, A. and {Natoli}, P. and {N{\o}rgaard-Nielsen}, H.~U. and {Pagano}, L. and {Paoletti}, D. and {Partridge}, B. and {Patanchon}, G. and {Peiris}, H.~V. and {Perrotta}, F. and {Pettorino}, V. and {Piacentini}, F. and {Polastri}, L. and {Polenta}, G. and {Puget}, J.-L. and {Rachen}, J.~P. and {Reinecke}, M. and {Remazeilles}, M. and {Renzi}, A. and {Rocha}, G. and {Rosset}, C. and {Roudier}, G. and {Rubi{\~n}o-Mart{\'\i}n}, J.~A. and {Ruiz-Granados}, B. and {Salvati}, L. and {Sandri}, M. and {Savelainen}, M. and {Scott}, D. and {Shellard}, E.~P.~S. and {Sirignano}, C. and {Sirri}, G. and {Spencer}, L.~D. and {Sunyaev}, R. and {Suur-Uski}, A.-S. and {Tauber}, J.~A. and {Tavagnacco}, D. and {Tenti}, M. and {Toffolatti}, L. and {Tomasi}, M. and {Trombetti}, T. and {Valenziano}, L. and {Valiviita}, J. and {Van Tent}, B. and {Vibert}, L. and {Vielva}, P. and {Villa}, F. and {Vittorio}, N. and {Wandelt}, B.~D. and {Wehus}, I.~K. and {White}, M. and {White}, S.~D.~M. and {Zacchei}, A. and {Zonca}, A.},
        title = "{Planck 2018 results. VI. Cosmological parameters}",
      journal = {\aap},
         year = 2020,
        month = sep,
       volume = {641},
          eid = {A6},
        pages = {A6},
          doi = {10.1051/0004-6361/201833910},
archivePrefix = {arXiv},
       eprint = {1807.06209},
 primaryClass = {astro-ph.CO},
       adsurl = {https://ui.adsabs.harvard.edu/abs/2020A&A...641A...6P}
}

@ARTICLE{Norris2006,
       author = {{Norris}, J.~P. and {Bonnell}, J.~T.},
        title = "{Short Gamma-Ray Bursts with Extended Emission}",
      journal = {\apj},
         year = 2006,
        month = may,
       volume = {643},
       number = {1},
        pages = {266-275},
          doi = {10.1086/502796},
archivePrefix = {arXiv},
       eprint = {astro-ph/0601190},
 primaryClass = {astro-ph},
       adsurl = {https://ui.adsabs.harvard.edu/abs/2006ApJ...643..266N}
}

@ARTICLE{Amati2002,
       author = {{Amati}, L. and {Frontera}, F. and {Tavani}, M. and {in't Zand}, J.~J.~M. and {Antonelli}, A. and {Costa}, E. and {Feroci}, M. and {Guidorzi}, C. and {Heise}, J. and {Masetti}, N. and {Montanari}, E. and {Nicastro}, L. and {Palazzi}, E. and {Pian}, E. and {Piro}, L. and {Soffitta}, P.},
        title = "{Intrinsic spectra and energetics of BeppoSAX Gamma-Ray Bursts with known redshifts}",
      journal = {\aap},
         year = 2002,
        month = jul,
       volume = {390},
        pages = {81-89},
          doi = {10.1051/0004-6361:20020722},
archivePrefix = {arXiv},
       eprint = {astro-ph/0205230},
 primaryClass = {astro-ph},
       adsurl = {https://ui.adsabs.harvard.edu/abs/2002A&A...390...81A}
}

@dataset{VISTADR5,
       author = {{McMahon}, R.~G. and {Banerji}, M. and {Gonzalez}, E. and {Koposov}, S.~E. and {Bejar}, V.~J. and {Lodieu}, N. and {Rebolo}, R. and {VHS Collaboration}},
        title = "{VizieR Online Data Catalog: The VISTA Hemisphere Survey (VHS) catalog DR5 (McMahon+, 2020)}",
 howpublished = {VizieR On-line Data Catalog: II/367.  Originally published in: 2013Msngr.154...35M},
         year = 2021,
        month = jan,
          eid = {II/367},
       adsurl = {https://ui.adsabs.harvard.edu/abs/2021yCat.2367....0M}
}

@ARTICLE{Windhorst2023,
       author = {{Windhorst}, Rogier A. and {Cohen}, Seth H. and {Jansen}, Rolf A. and {Summers}, Jake and {Tompkins}, Scott and {Conselice}, Christopher J. and {Driver}, Simon P. and {Yan}, Haojing and {Coe}, Dan and {Frye}, Brenda and {Grogin}, Norman and {Koekemoer}, Anton and {Marshall}, Madeline A. and {O'Brien}, Rosalia and {Pirzkal}, Nor and {Robotham}, Aaron and {Ryan}, Russell E. and {Willmer}, Christopher N.~A. and {Carleton}, Timothy and {Diego}, Jose M. and {Keel}, William C. and {Porto}, Paolo and {Redshaw}, Caleb and {Scheller}, Sydney and {Wilkins}, Stephen M. and {Willner}, S.~P. and {Zitrin}, Adi and {Adams}, Nathan J. and {Austin}, Duncan and {Arendt}, Richard G. and {Beacom}, John F. and {Bhatawdekar}, Rachana A. and {Bradley}, Larry D. and {Broadhurst}, Tom and {Cheng}, Cheng and {Civano}, Francesca and {Dai}, Liang and {Dole}, Herv{\'e} and {D'Silva}, Jordan C.~J. and {Duncan}, Kenneth J. and {Fazio}, Giovanni G. and {Ferrami}, Giovanni and {Ferreira}, Leonardo and {Finkelstein}, Steven L. and {Furtak}, Lukas J. and {Gim}, Hansung B. and {Griffiths}, Alex and {Hammel}, Heidi B. and {Harrington}, Kevin C. and {Hathi}, Nimish P. and {Holwerda}, Benne W. and {Honor}, Rachel and {Huang}, Jia-Sheng and {Hyun}, Minhee and {Im}, Myungshin and {Joshi}, Bhavin A. and {Kamieneski}, Patrick S. and {Kelly}, Patrick and {Larson}, Rebecca L. and {Li}, Juno and {Lim}, Jeremy and {Ma}, Zhiyuan and {Maksym}, Peter and {Manzoni}, Giorgio and {Meena}, Ashish Kumar and {Milam}, Stefanie N. and {Nonino}, Mario and {Pascale}, Massimo and {Petric}, Andreea and {Pierel}, Justin D.~R. and {Polletta}, Maria del Carmen and {R{\"o}ttgering}, Huub J.~A. and {Rutkowski}, Michael J. and {Smail}, Ian and {Straughn}, Amber N. and {Strolger}, Louis-Gregory and {Swirbul}, Andi and {Trussler}, James A.~A. and {Wang}, Lifan and {Welch}, Brian and {B. Wyithe}, J. Stuart and {Yun}, Min and {Zackrisson}, Erik and {Zhang}, Jiashuo and {Zhao}, Xiurui},
        title = "{JWST PEARLS. Prime Extragalactic Areas for Reionization and Lensing Science: Project Overview and First Results}",
      journal = {\aj},
         year = 2023,
        month = jan,
       volume = {165},
       number = {1},
          eid = {13},
        pages = {13},
          doi = {10.3847/1538-3881/aca163},
archivePrefix = {arXiv},
       eprint = {2209.04119},
 primaryClass = {astro-ph.CO},
       adsurl = {https://ui.adsabs.harvard.edu/abs/2023AJ....165...13W}
}

@ARTICLE{Leibler2010,
       author = {{Leibler}, C.~N. and {Berger}, E.},
        title = "{The Stellar Ages and Masses of Short Gamma-ray Burst Host Galaxies: Investigating the Progenitor Delay Time Distribution and the Role of Mass and Star Formation in the Short Gamma-ray Burst Rate}",
      journal = {\apj},
         year = 2010,
        month = dec,
       volume = {725},
       number = {1},
        pages = {1202-1214},
          doi = {10.1088/0004-637X/725/1/1202},
archivePrefix = {arXiv},
       eprint = {1009.1147},
 primaryClass = {astro-ph.HE},
       adsurl = {https://ui.adsabs.harvard.edu/abs/2010ApJ...725.1202L}
}

@ARTICLE{Troja2023,
       author = {{Troja}, Eleonora},
        title = "{Eighteen Years of Kilonova Discoveries with Swift}",
      journal = {Universe},
         year = 2023,
        month = may,
       volume = {9},
       number = {6},
          eid = {245},
        pages = {245},
          doi = {10.3390/universe9060245},
archivePrefix = {arXiv},
       eprint = {2305.18531},
 primaryClass = {astro-ph.HE},
       adsurl = {https://ui.adsabs.harvard.edu/abs/2023Univ....9..245T}
}

@ARTICLE{Kumar2000,
       author = {{Kumar}, Pawan and {Panaitescu}, Alin},
        title = "{Afterglow Emission from Naked Gamma-Ray Bursts}",
      journal = {\apjl},
         year = 2000,
        month = oct,
       volume = {541},
       number = {2},
        pages = {L51-L54},
          doi = {10.1086/312905},
archivePrefix = {arXiv},
       eprint = {astro-ph/0006317},
 primaryClass = {astro-ph},
       adsurl = {https://ui.adsabs.harvard.edu/abs/2000ApJ...541L..51K}
}

@ARTICLE{OConnor2020,
       author = {{O'Connor}, Brendan and {Beniamini}, Paz and {Kouveliotou}, Chryssa},
        title = "{Constraints on the circumburst environments of short gamma-ray bursts}",
      journal = {\mnras},
         year = 2020,
        month = jul,
       volume = {495},
       number = {4},
        pages = {4782-4799},
          doi = {10.1093/mnras/staa1433},
archivePrefix = {arXiv},
       eprint = {2004.00031},
 primaryClass = {astro-ph.HE},
       adsurl = {https://ui.adsabs.harvard.edu/abs/2020MNRAS.495.4782O}
}

@ARTICLE{Yang2024,
       author = {{Yang}, Yu-Han and {Troja}, Eleonora and {O'Connor}, Brendan and {Fryer}, Chris L. and {Im}, Myungshin and {Durbak}, Joe and {Paek}, Gregory S.~H. and {Ricci}, Roberto and {Bom}, Cl{\'e}cio R. and {Gillanders}, James H. and {Castro-Tirado}, Alberto J. and {Peng}, Zong-Kai and {Dichiara}, Simone and {Ryan}, Geoffrey and {van Eerten}, Hendrik and {Dai}, Zi-Gao and {Chang}, Seo-Won and {Choi}, Hyeonho and {De}, Kishalay and {Hu}, Youdong and {Kilpatrick}, Charles D. and {Kutyrev}, Alexander and {Jeong}, Mankeun and {Lee}, Chung-Uk and {Makler}, Martin and {Navarete}, Felipe and {P{\'e}rez-Garc{\'\i}a}, Ignacio},
        title = "{A lanthanide-rich kilonova in the aftermath of a long gamma-ray burst}",
      journal = {\nat},
         year = 2024,
        month = feb,
       volume = {626},
       number = {8000},
        pages = {742-745},
          doi = {10.1038/s41586-023-06979-5},
archivePrefix = {arXiv},
       eprint = {2308.00638},
 primaryClass = {astro-ph.HE},
       adsurl = {https://ui.adsabs.harvard.edu/abs/2024Natur.626..742Y}
}

@ARTICLE{Barthelmy2004,
       author = {{Barthelmy}, Scott D. and {Barbier}, Louis M. and {Cummings}, Jay R. and {Fenimore}, Ed E. and {Gehrels}, Neil and {Hullinger}, Derek and {Krimm}, Hans A. and {Markwardt}, Craig B. and {Palmer}, David M. and {Parsons}, Ann and {Sato}, Goro and {Suzuki}, Masaya and {Takahashi}, Tadayuki and {Tashiro}, Makota and {Tueller}, Jack},
        title = "{The Burst Alert Telescope (BAT) on the SWIFT Midex Mission}",
      journal = {\ssr},
         year = 2005,
        month = oct,
       volume = {120},
       number = {3-4},
        pages = {143-164},
          doi = {10.1007/s11214-005-5096-3},
archivePrefix = {arXiv},
       eprint = {astro-ph/0507410},
 primaryClass = {astro-ph},
       adsurl = {https://ui.adsabs.harvard.edu/abs/2005SSRv..120..143B}
}

@ARTICLE{Dichiara2026,
       author = {{Dichiara}, S. and {Troja}, E. and {O'Connor}, B. and {Yang}, Y.-H. and {Beniamini}, P. and {Galvan-Gamez}, A. and {Sakamoto}, T. and {Kawakubo}, Y. and {Charlton}, J.~C.},
        title = "{A Merger within a Merger: Chandra Pinpoints the Short GRB 230906A in a Peculiar Environment}",
      journal = {\apjl},
         year = 2026,
        month = mar,
       volume = {999},
       number = {2},
          eid = {L42},
        pages = {L42},
          doi = {10.3847/2041-8213/ae2a2f},
archivePrefix = {arXiv},
       eprint = {2510.15867},
 primaryClass = {astro-ph.HE},
       adsurl = {https://ui.adsabs.harvard.edu/abs/2026ApJ...999L..42D}
}

@ARTICLE{Roming2005,
       author = {{Roming}, Peter W.~A. and {Kennedy}, Thomas E. and {Mason}, Keith O. and {Nousek}, John A. and {Ahr}, Lindy and {Bingham}, Richard E. and {Broos}, Patrick S. and {Carter}, Mary J. and {Hancock}, Barry K. and {Huckle}, Howard E. and {Hunsberger}, S.~D. and {Kawakami}, Hajime and {Killough}, Ronnie and {Koch}, T. Scott and {McLelland}, Michael K. and {Smith}, Kelly and {Smith}, Philip J. and {Soto}, Juan Carlos and {Boyd}, Patricia T. and {Breeveld}, Alice A. and {Holland}, Stephen T. and {Ivanushkina}, Mariya and {Pryzby}, Michael S. and {Still}, Martin D. and {Stock}, Joseph},
        title = "{The Swift Ultra-Violet/Optical Telescope}",
      journal = {\ssr},
         year = 2005,
        month = oct,
       volume = {120},
       number = {3-4},
        pages = {95-142},
          doi = {10.1007/s11214-005-5095-4},
archivePrefix = {arXiv},
       eprint = {astro-ph/0507413},
 primaryClass = {astro-ph},
       adsurl = {https://ui.adsabs.harvard.edu/abs/2005SSRv..120...95R}
}

@ARTICLE{Troja2022,
       author = {{Troja}, E. and {Fryer}, C.~L. and {O'Connor}, B. and {Ryan}, G. and {Dichiara}, S. and {Kumar}, A. and {Ito}, N. and {Gupta}, R. and {Wollaeger}, R.~T. and {Norris}, J.~P. and {Kawai}, N. and {Butler}, N.~R. and {Aryan}, A. and {Misra}, K. and {Hosokawa}, R. and {Murata}, K.~L. and {Niwano}, M. and {Pandey}, S.~B. and {Kutyrev}, A. and {van Eerten}, H.~J. and {Chase}, E.~A. and {Hu}, Y.-D. and {Caballero-Garcia}, M.~D. and {Castro-Tirado}, A.~J.},
        title = "{A nearby long gamma-ray burst from a merger of compact objects}",
      journal = {\nat},
         year = 2022,
        month = dec,
       volume = {612},
       number = {7939},
        pages = {228-231},
          doi = {10.1038/s41586-022-05327-3},
archivePrefix = {arXiv},
       eprint = {2209.03363},
 primaryClass = {astro-ph.HE},
       adsurl = {https://ui.adsabs.harvard.edu/abs/2022Natur.612..228T}
}

@ARTICLE{Mao2026,
       author = {{Mao}, Yuhan and {He}, Hanrui and {Ren}, Jia and {Wang}, Yun and {Zhou}, Hao and {Wang}, Qiuli and {Zhu}, Yiming and {Jin}, Zhiping and {Wei}, Daming},
        title = "{Revealing the high redshift host galaxy of the short GRB 061201 with JWST}",
      journal = {arXiv e-prints},
         year = 2026,
        month = jun,
          eid = {arXiv:2606.02032},
        pages = {arXiv:2606.02032},
          doi = {10.48550/arXiv.2606.02032},
archivePrefix = {arXiv},
       eprint = {2606.02032},
 primaryClass = {astro-ph.HE},
       adsurl = {https://ui.adsabs.harvard.edu/abs/2026arXiv260602032M}
}

@ARTICLE{Fong2013,
       author = {{Fong}, W. and {Berger}, E.},
        title = "{The Locations of Short Gamma-Ray Bursts as Evidence for Compact Object Binary Progenitors}",
      journal = {\apj},
         year = 2013,
        month = oct,
       volume = {776},
       number = {1},
          eid = {18},
        pages = {18},
          doi = {10.1088/0004-637X/776/1/18},
archivePrefix = {arXiv},
       eprint = {1307.0819},
 primaryClass = {astro-ph.HE},
       adsurl = {https://ui.adsabs.harvard.edu/abs/2013ApJ...776...18F}
}

@ARTICLE{Abbott2017a,
       author = {{Abbott}, B.~P. and {Abbott}, R. and {Abbott}, T.~D. and {Acernese}, F. and {Ackley}, K. and {Adams}, C. and {Adams}, T. and {Addesso}, P. and {Adhikari}, R.~X. and {Adya}, V.~B. and {Affeldt}, C. and {Afrough}, M. and {Agarwal}, B. and {Agathos}, M. and {Agatsuma}, K. and {Aggarwal}, N. and {Aguiar}, O.~D. and {Aiello}, L. and {Ain}, A. and {Ajith}, P. and {Allen}, B. and {Allen}, G. and {Allocca}, A. and {Aloy}, M.~A. and {Altin}, P.~A. and {Amato}, A. and {Ananyeva}, A. and {Anderson}, S.~B. and {Anderson}, W.~G. and {Angelova}, S.~V. and {Antier}, S. and {Appert}, S. and {Arai}, K. and {Araya}, M.~C. and {Areeda}, J.~S. and {Arnaud}, N. and {Arun}, K.~G. and {Ascenzi}, S. and {Ashton}, G. and {Ast}, M. and {Aston}, S.~M. and {Astone}, P. and {Atallah}, D.~V. and {Aufmuth}, P. and {Aulbert}, C. and {AultONeal}, K. and {Austin}, C. and {Avila-Alvarez}, A. and {Babak}, S. and {Bacon}, P. and {Bader}, M.~K.~M. and {Bae}, S. and {Baker}, P.~T. and {Baldaccini}, F. and {Ballardin}, G. and {Ballmer}, S.~W. and {Banagiri}, S. and {Barayoga}, J.~C. and {Barclay}, S.~E. and {Barish}, B.~C. and {Barker}, D. and {Barkett}, K. and {Barone}, F. and {Barr}, B. and {Barsotti}, L. and {Barsuglia}, M. and {Barta}, D. and {Bartlett}, J. and {Bartos}, I. and {Bassiri}, R. and {Basti}, A. and {Batch}, J.~C. and {Bawaj}, M. and {Bayley}, J.~C. and {Bazzan}, M. and {B{\'e}csy}, B. and {Beer}, C. and {Bejger}, M. and {Belahcene}, I. and {Bell}, A.~S. and {Berger}, B.~K. and {Bergmann}, G. and {Bero}, J.~J. and {Berry}, C.~P.~L. and {Bersanetti}, D. and {Bertolini}, A. and {Betzwieser}, J. and {Bhagwat}, S. and {Bhandare}, R. and {Bilenko}, I.~A. and {Billingsley}, G. and {Billman}, C.~R. and {Birch}, J. and {Birney}, R. and {Birnholtz}, O. and {Biscans}, S. and {Biscoveanu}, S. and {Bisht}, A. and {Bitossi}, M. and {Biwer}, C. and {Bizouard}, M.~A. and {Blackburn}, J.~K. and {Blackman}, J. and {Blair}, C.~D. and {Blair}, D.~G. and {Blair}, R.~M. and {Bloemen}, S. and {Bock}, O. and {Bode}, N. and {Boer}, M. and {Bogaert}, G. and {Bohe}, A. and {Bondu}, F. and {Bonilla}, E. and {Bonnand}, R. and {Boom}, B.~A. and {Bork}, R. and {Boschi}, V. and {Bose}, S. and {Bossie}, K. and {Bouffanais}, Y. and {Bozzi}, A. and {Bradaschia}, C. and {Brady}, P.~R. and {Branchesi}, M. and {Brau}, J.~E. and {Briant}, T. and {Brillet}, A. and {Brinkmann}, M. and {Brisson}, V. and {Brockill}, P. and {Broida}, J.~E. and {Brooks}, A.~F. and {Brown}, D.~A. and {Brown}, D.~D. and {Brunett}, S. and {Buchanan}, C.~C. and {Buikema}, A. and {Bulik}, T. and {Bulten}, H.~J. and {Buonanno}, A. and {Buskulic}, D. and {Buy}, C. and {Byer}, R.~L. and {Cabero}, M. and {Cadonati}, L. and {Cagnoli}, G. and {Cahillane}, C. and {Calder{\'o}n Bustillo}, J. and {Callister}, T.~A. and {Calloni}, E. and {Camp}, J.~B. and {Canepa}, M. and {Canizares}, P. and {Cannon}, K.~C. and {Cao}, H. and {Cao}, J. and {Capano}, C.~D. and {Capocasa}, E. and {Carbognani}, F. and {Caride}, S. and {Carney}, M.~F. and {Casanueva Diaz}, J. and {Casentini}, C. and {Caudill}, S. and {Cavagli{\`a}}, M. and {Cavalier}, F. and {Cavalieri}, R. and {Cella}, G. and {Cepeda}, C.~B. and {Cerd{\'a}-Dur{\'a}n}, P. and {Cerretani}, G. and {Cesarini}, E. and {Chamberlin}, S.~J. and {Chan}, M. and {Chao}, S. and {Charlton}, P. and {Chase}, E. and {Chassande-Mottin}, E. and {Chatterjee}, D. and {Chatziioannou}, K. and {Cheeseboro}, B.~D. and {Chen}, H.~Y. and {Chen}, X. and {Chen}, Y. and {Cheng}, H.-P. and {Chia}, H. and {Chincarini}, A. and {Chiummo}, A. and {Chmiel}, T. and {Cho}, H.~S. and {Cho}, M. and {Chow}, J.~H. and {Christensen}, N. and {Chu}, Q. and {Chua}, A.~J.~K. and {Chua}, S. and {Chung}, A.~K.~W. and {Chung}, S. and {Ciani}, G.},
        title = "{Gravitational Waves and Gamma-Rays from a Binary Neutron Star Merger: GW170817 and GRB 170817A}",
      journal = {\apjl},
         year = 2017,
        month = oct,
       volume = {848},
       number = {2},
          eid = {L13},
        pages = {L13},
          doi = {10.3847/2041-8213/aa920c},
archivePrefix = {arXiv},
       eprint = {1710.05834},
 primaryClass = {astro-ph.HE},
       adsurl = {https://ui.adsabs.harvard.edu/abs/2017ApJ...848L..13A}
}

@ARTICLE{Abbott17b,
       author = {{Abbott}, B.~P. and {Abbott}, R. and {Abbott}, T.~D. and {Acernese}, F. and {Ackley}, K. and {Adams}, C. and {Adams}, T. and {Addesso}, P. and {Adhikari}, R.~X. and {Adya}, V.~B. and {Affeldt}, C. and {Afrough}, M. and {Agarwal}, B. and {Agathos}, M. and {Agatsuma}, K. and {Aggarwal}, N. and {Aguiar}, O.~D. and {Aiello}, L. and {Ain}, A. and {Ajith}, P. and {Allen}, B. and {Allen}, G. and {Allocca}, A. and {Altin}, P.~A. and {Amato}, A. and {Ananyeva}, A. and {Anderson}, S.~B. and {Anderson}, W.~G. and {Angelova}, S.~V. and {Antier}, S. and {Appert}, S. and {Arai}, K. and {Araya}, M.~C. and {Areeda}, J.~S. and {Arnaud}, N. and {Arun}, K.~G. and {Ascenzi}, S. and {Ashton}, G. and {Ast}, M. and {Aston}, S.~M. and {Astone}, P. and {Atallah}, D.~V. and {Aufmuth}, P. and {Aulbert}, C. and {AultONeal}, K. and {Austin}, C. and {Avila-Alvarez}, A. and {Babak}, S. and {Bacon}, P. and {Bader}, M.~K.~M. and {Bae}, S. and {Bailes}, M. and {Baker}, P.~T. and {Baldaccini}, F. and {Ballardin}, G. and {Ballmer}, S.~W. and {Banagiri}, S. and {Barayoga}, J.~C. and {Barclay}, S.~E. and {Barish}, B.~C. and {Barker}, D. and {Barkett}, K. and {Barone}, F. and {Barr}, B. and {Barsotti}, L. and {Barsuglia}, M. and {Barta}, D. and {Barthelmy}, S.~D. and {Bartlett}, J. and {Bartos}, I. and {Bassiri}, R. and {Basti}, A. and {Batch}, J.~C. and {Bawaj}, M. and {Bayley}, J.~C. and {Bazzan}, M. and {B{\'e}csy}, B. and {Beer}, C. and {Bejger}, M. and {Belahcene}, I. and {Bell}, A.~S. and {Berger}, B.~K. and {Bergmann}, G. and {Bernuzzi}, S. and {Bero}, J.~J. and {Berry}, C.~P.~L. and {Bersanetti}, D. and {Bertolini}, A. and {Betzwieser}, J. and {Bhagwat}, S. and {Bhandare}, R. and {Bilenko}, I.~A. and {Billingsley}, G. and {Billman}, C.~R. and {Birch}, J. and {Birney}, R. and {Birnholtz}, O. and {Biscans}, S. and {Biscoveanu}, S. and {Bisht}, A. and {Bitossi}, M. and {Biwer}, C. and {Bizouard}, M.~A. and {Blackburn}, J.~K. and {Blackman}, J. and {Blair}, C.~D. and {Blair}, D.~G. and {Blair}, R.~M. and {Bloemen}, S. and {Bock}, O. and {Bode}, N. and {Boer}, M. and {Bogaert}, G. and {Bohe}, A. and {Bondu}, F. and {Bonilla}, E. and {Bonnand}, R. and {Boom}, B.~A. and {Bork}, R. and {Boschi}, V. and {Bose}, S. and {Bossie}, K. and {Bouffanais}, Y. and {Bozzi}, A. and {Bradaschia}, C. and {Brady}, P.~R. and {Branchesi}, M. and {Brau}, J.~E. and {Briant}, T. and {Brillet}, A. and {Brinkmann}, M. and {Brisson}, V. and {Brockill}, P. and {Broida}, J.~E. and {Brooks}, A.~F. and {Brown}, D.~A. and {Brown}, D.~D. and {Brunett}, S. and {Buchanan}, C.~C. and {Buikema}, A. and {Bulik}, T. and {Bulten}, H.~J. and {Buonanno}, A. and {Buskulic}, D. and {Buy}, C. and {Byer}, R.~L. and {Cabero}, M. and {Cadonati}, L. and {Cagnoli}, G. and {Cahillane}, C. and {Calder{\'o}n Bustillo}, J. and {Callister}, T.~A. and {Calloni}, E. and {Camp}, J.~B. and {Canepa}, M. and {Canizares}, P. and {Cannon}, K.~C. and {Cao}, H. and {Cao}, J. and {Capano}, C.~D. and {Capocasa}, E. and {Carbognani}, F. and {Caride}, S. and {Carney}, M.~F. and {Carullo}, G. and {Casanueva Diaz}, J. and {Casentini}, C. and {Caudill}, S. and {Cavagli{\`a}}, M. and {Cavalier}, F. and {Cavalieri}, R. and {Cella}, G. and {Cepeda}, C.~B. and {Cerd{\'a}-Dur{\'a}n}, P. and {Cerretani}, G. and {Cesarini}, E. and {Chamberlin}, S.~J. and {Chan}, M. and {Chao}, S. and {Charlton}, P. and {Chase}, E. and {Chassande-Mottin}, E. and {Chatterjee}, D. and {Chatziioannou}, K. and {Cheeseboro}, B.~D. and {Chen}, H.~Y. and {Chen}, X. and {Chen}, Y. and {Cheng}, H.-P. and {Chia}, H. and {Chincarini}, A. and {Chiummo}, A. and {Chmiel}, T. and {Cho}, H.~S. and {Cho}, M. and {Chow}, J.~H. and {Christensen}, N. and {Chu}, Q. and {Chua}, A.~J.~K. and {Chua}, S.},
        title = "{GW170817: Observation of Gravitational Waves from a Binary Neutron Star Inspiral}",
      journal = {\prl},
         year = 2017,
        month = oct,
       volume = {119},
       number = {16},
          eid = {161101},
        pages = {161101},
          doi = {10.1103/PhysRevLett.119.161101},
archivePrefix = {arXiv},
       eprint = {1710.05832},
 primaryClass = {gr-qc},
       adsurl = {https://ui.adsabs.harvard.edu/abs/2017PhRvL.119p1101A}
}

@ARTICLE{Hotokezaka2023,
       author = {{Hotokezaka}, Kenta and {Tanaka}, Masaomi and {Kato}, Daiji and {Gaigalas}, Gediminas},
        title = "{Tellurium emission line in kilonova AT 2017gfo}",
      journal = {\mnras},
         year = 2023,
        month = nov,
       volume = {526},
       number = {1},
        pages = {L155-L159},
          doi = {10.1093/mnrasl/slad128},
archivePrefix = {arXiv},
       eprint = {2307.00988},
 primaryClass = {astro-ph.HE},
       adsurl = {https://ui.adsabs.harvard.edu/abs/2023MNRAS.526L.155H}
}

@ARTICLE{Willingale2013,
       author = {{Willingale}, R. and {Starling}, R.~L.~C. and {Beardmore}, A.~P. and {Tanvir}, N.~R. and {O'Brien}, P.~T.},
        title = "{Calibration of X-ray absorption in our Galaxy}",
      journal = {\mnras},
         year = 2013,
        month = may,
       volume = {431},
       number = {1},
        pages = {394-404},
          doi = {10.1093/mnras/stt175},
archivePrefix = {arXiv},
       eprint = {1303.0843},
 primaryClass = {astro-ph.HE},
       adsurl = {https://ui.adsabs.harvard.edu/abs/2013MNRAS.431..394W}
}

@ARTICLE{SF2011,
       author = {{Schlafly}, Edward F. and {Finkbeiner}, Douglas P.},
        title = "{Measuring Reddening with Sloan Digital Sky Survey Stellar Spectra and Recalibrating SFD}",
      journal = {\apj},
         year = 2011,
        month = aug,
       volume = {737},
       number = {2},
          eid = {103},
        pages = {103},
          doi = {10.1088/0004-637X/737/2/103},
archivePrefix = {arXiv},
       eprint = {1012.4804},
 primaryClass = {astro-ph.GA},
       adsurl = {https://ui.adsabs.harvard.edu/abs/2011ApJ...737..103S}
}

@ARTICLE{Kasen2017,
       author = {{Kasen}, Daniel and {Metzger}, Brian and {Barnes}, Jennifer and {Quataert}, Eliot and {Ramirez-Ruiz}, Enrico},
        title = "{Origin of the heavy elements in binary neutron-star mergers from a gravitational-wave event}",
      journal = {\nat},
         year = 2017,
        month = nov,
       volume = {551},
       number = {7678},
        pages = {80-84},
          doi = {10.1038/nature24453},
archivePrefix = {arXiv},
       eprint = {1710.05463},
 primaryClass = {astro-ph.HE},
       adsurl = {https://ui.adsabs.harvard.edu/abs/2017Natur.551...80K}
}

@ARTICLE{Pian2017,
       author = {{Pian}, E. and {D'Avanzo}, P. and {Benetti}, S. and {Branchesi}, M. and {Brocato}, E. and {Campana}, S. and {Cappellaro}, E. and {Covino}, S. and {D'Elia}, V. and {Fynbo}, J.~P.~U. and {Getman}, F. and {Ghirlanda}, G. and {Ghisellini}, G. and {Grado}, A. and {Greco}, G. and {Hjorth}, J. and {Kouveliotou}, C. and {Levan}, A. and {Limatola}, L. and {Malesani}, D. and {Mazzali}, P.~A. and {Melandri}, A. and {M{\o}ller}, P. and {Nicastro}, L. and {Palazzi}, E. and {Piranomonte}, S. and {Rossi}, A. and {Salafia}, O.~S. and {Selsing}, J. and {Stratta}, G. and {Tanaka}, M. and {Tanvir}, N.~R. and {Tomasella}, L. and {Watson}, D. and {Yang}, S. and {Amati}, L. and {Antonelli}, L.~A. and {Ascenzi}, S. and {Bernardini}, M.~G. and {Bo{\"e}r}, M. and {Bufano}, F. and {Bulgarelli}, A. and {Capaccioli}, M. and {Casella}, P. and {Castro-Tirado}, A.~J. and {Chassande-Mottin}, E. and {Ciolfi}, R. and {Copperwheat}, C.~M. and {Dadina}, M. and {De Cesare}, G. and {di Paola}, A. and {Fan}, Y.~Z. and {Gendre}, B. and {Giuffrida}, G. and {Giunta}, A. and {Hunt}, L.~K. and {Israel}, G.~L. and {Jin}, Z.-P. and {Kasliwal}, M.~M. and {Klose}, S. and {Lisi}, M. and {Longo}, F. and {Maiorano}, E. and {Mapelli}, M. and {Masetti}, N. and {Nava}, L. and {Patricelli}, B. and {Perley}, D. and {Pescalli}, A. and {Piran}, T. and {Possenti}, A. and {Pulone}, L. and {Razzano}, M. and {Salvaterra}, R. and {Schipani}, P. and {Spera}, M. and {Stamerra}, A. and {Stella}, L. and {Tagliaferri}, G. and {Testa}, V. and {Troja}, E. and {Turatto}, M. and {Vergani}, S.~D. and {Vergani}, D.},
        title = "{Spectroscopic identification of r-process nucleosynthesis in a double neutron-star merger}",
      journal = {\nat},
         year = 2017,
        month = nov,
       volume = {551},
       number = {7678},
        pages = {67-70},
          doi = {10.1038/nature24298},
archivePrefix = {arXiv},
       eprint = {1710.05858},
 primaryClass = {astro-ph.HE},
       adsurl = {https://ui.adsabs.harvard.edu/abs/2017Natur.551...67P}
}

@ARTICLE{Eichler89,
       author = {{Eichler}, David and {Livio}, Mario and {Piran}, Tsvi and {Schramm}, David N.},
        title = "{Nucleosynthesis, neutrino bursts and {\ensuremath{\gamma}}-rays from coalescing neutron stars}",
      journal = {\nat},
         year = 1989,
        month = jul,
       volume = {340},
       number = {6229},
        pages = {126-128},
          doi = {10.1038/340126a0},
       adsurl = {https://ui.adsabs.harvard.edu/abs/1989Natur.340..126E}
}

@ARTICLE{Moss2022,
       author = {{Moss}, Michael and {Lien}, Amy and {Guiriec}, Sylvain and {Cenko}, S. Bradley and {Sakamoto}, Takanori},
        title = "{Instrumental Tip-of-the-iceberg Effects on the Prompt Emission of Swift/BAT Gamma-ray Bursts}",
      journal = {\apj},
         year = 2022,
        month = mar,
       volume = {927},
       number = {2},
          eid = {157},
        pages = {157},
          doi = {10.3847/1538-4357/ac4d94},
archivePrefix = {arXiv},
       eprint = {2111.13392},
 primaryClass = {astro-ph.HE},
       adsurl = {https://ui.adsabs.harvard.edu/abs/2022ApJ...927..157M}
}

@ARTICLE{Piran1992,
       author = {{Piran}, Tsvi},
        title = "{The Implications of the Compton (GRO) Observations for Cosmological Gamma-Ray Bursts}",
      journal = {\apjl},
         year = 1992,
        month = apr,
       volume = {389},
        pages = {L45},
          doi = {10.1086/186345},
       adsurl = {https://ui.adsabs.harvard.edu/abs/1992ApJ...389L..45P}
}

@ARTICLE{Wanderman2015,
       author = {{Wanderman}, David and {Piran}, Tsvi},
        title = "{The rate, luminosity function and time delay of non-Collapsar short GRBs}",
      journal = {\mnras},
         year = 2015,
        month = apr,
       volume = {448},
       number = {4},
        pages = {3026-3037},
          doi = {10.1093/mnras/stv123},
archivePrefix = {arXiv},
       eprint = {1405.5878},
 primaryClass = {astro-ph.HE},
       adsurl = {https://ui.adsabs.harvard.edu/abs/2015MNRAS.448.3026W}
}

@ARTICLE{MadauFragos2017,
       author = {{Madau}, Piero and {Fragos}, Tassos},
        title = "{Radiation Backgrounds at Cosmic Dawn: X-Rays from Compact Binaries}",
      journal = {\apj},
         year = 2017,
        month = may,
       volume = {840},
       number = {1},
          eid = {39},
        pages = {39},
          doi = {10.3847/1538-4357/aa6af9},
archivePrefix = {arXiv},
       eprint = {1606.07887},
 primaryClass = {astro-ph.GA},
       adsurl = {https://ui.adsabs.harvard.edu/abs/2017ApJ...840...39M}
}

@ARTICLE{Levesque2010,
       author = {{Levesque}, Emily M. and {Bloom}, Joshua S. and {Butler}, Nathaniel R. and {Perley}, Daniel A. and {Cenko}, S. Bradley and {Prochaska}, J. Xavier and {Kewley}, Lisa J. and {Bunker}, Andrew and {Chen}, Hsiao-Wen and {Chornock}, Ryan and {Filippenko}, Alexei V. and {Glazebrook}, Karl and {Lopez}, Sebastian and {Masiero}, Joseph and {Modjaz}, Maryam and {Morgan}, Adam and {Poznanski}, Dovi},
        title = "{GRB090426: the environment of a rest-frame 0.35-s gamma-ray burst at a redshift of 2.609}",
      journal = {\mnras},
         year = 2010,
        month = jan,
       volume = {401},
       number = {2},
        pages = {963-972},
          doi = {10.1111/j.1365-2966.2009.15733.x},
archivePrefix = {arXiv},
       eprint = {0907.1661},
 primaryClass = {astro-ph.HE},
       adsurl = {https://ui.adsabs.harvard.edu/abs/2010MNRAS.401..963L}
}

@ARTICLE{Bromberg2012,
       author = {{Bromberg}, Omer and {Nakar}, Ehud and {Piran}, Tsvi and {Sari}, Re'em},
        title = "{An Observational Imprint of the Collapsar Model of Long Gamma-Ray Bursts}",
      journal = {\apj},
         year = 2012,
        month = apr,
       volume = {749},
       number = {2},
          eid = {110},
        pages = {110},
          doi = {10.1088/0004-637X/749/2/110},
archivePrefix = {arXiv},
       eprint = {1111.2990},
 primaryClass = {astro-ph.HE},
       adsurl = {https://ui.adsabs.harvard.edu/abs/2012ApJ...749..110B}
}

@ARTICLE{Nugent2022,
       author = {{Nugent}, Anya E. and {Fong}, Wen-Fai and {Dong}, Yuxin and {Leja}, Joel and {Berger}, Edo and {Zevin}, Michael and {Chornock}, Ryan and {Cobb}, Bethany E. and {Kelley}, Luke Zoltan and {Kilpatrick}, Charles D. and {Levan}, Andrew and {Margutti}, Raffaella and {Paterson}, Kerry and {Perley}, Daniel and {Escorial}, Alicia Rouco and {Smith}, Nathan and {Tanvir}, Nial},
        title = "{Short GRB Host Galaxies. II. A Legacy Sample of Redshifts, Stellar Population Properties, and Implications for Their Neutron Star Merger Origins}",
      journal = {\apj},
         year = 2022,
        month = nov,
       volume = {940},
       number = {1},
          eid = {57},
        pages = {57},
          doi = {10.3847/1538-4357/ac91d1},
archivePrefix = {arXiv},
       eprint = {2206.01764},
 primaryClass = {astro-ph.GA},
       adsurl = {https://ui.adsabs.harvard.edu/abs/2022ApJ...940...57N}
}

@ARTICLE{Dichiara2021,
       author = {{Dichiara}, S. and {Troja}, E. and {Beniamini}, P. and {O'Connor}, B. and {Moss}, M. and {Lien}, A.~Y. and {Ricci}, R. and {Amati}, L. and {Ryan}, G. and {Sakamoto}, T.},
        title = "{Evidence of Extended Emission in GRB 181123B and Other High-redshift Short GRBs}",
      journal = {\apjl},
         year = 2021,
        month = apr,
       volume = {911},
       number = {2},
          eid = {L28},
        pages = {L28},
          doi = {10.3847/2041-8213/abf562},
archivePrefix = {arXiv},
       eprint = {2103.02558},
 primaryClass = {astro-ph.HE},
       adsurl = {https://ui.adsabs.harvard.edu/abs/2021ApJ...911L..28D}
}

@ARTICLE{Sakamoto2013,
       author = {{Sakamoto}, T. and {Troja}, E. and {Aoki}, K. and {Guiriec}, S. and {Im}, M. and {Leloudas}, G. and {Malesani}, D. and {Melandri}, A. and {de Ugarte Postigo}, A. and {Urata}, Y. and {Xu}, D. and {D'Avanzo}, P. and {Gorosabel}, J. and {Jeon}, Y. and {S{\'a}nchez-Ram{\'\i}rez}, R. and {Andersen}, M.~I. and {Bai}, J. and {Barthelmy}, S.~D. and {Briggs}, M.~S. and {Foley}, S. and {Fruchter}, A.~S. and {Fynbo}, J.~P.~U. and {Gehrels}, N. and {Huang}, K. and {Jang}, M. and {Kawai}, N. and {Korhonen}, H. and {Mao}, J. and {Norris}, J.~P. and {Preece}, R.~D. and {Racusin}, J.~L. and {Th{\"o}ne}, C.~C. and {Vida}, K. and {Zhao}, X.},
        title = "{Identifying the Location in the Host Galaxy of the Short GRB 111117A with the Chandra Subarcsecond Position}",
      journal = {\apj},
         year = 2013,
        month = mar,
       volume = {766},
       number = {1},
          eid = {41},
        pages = {41},
          doi = {10.1088/0004-637X/766/1/41},
archivePrefix = {arXiv},
       eprint = {1205.6774},
 primaryClass = {astro-ph.HE},
       adsurl = {https://ui.adsabs.harvard.edu/abs/2013ApJ...766...41S}
}

@ARTICLE{Belczynski2006,
       author = {{Belczynski}, Krzysztof and {Perna}, Rosalba and {Bulik}, Tomasz and {Kalogera}, Vassiliki and {Ivanova}, Natalia and {Lamb}, Donald Q.},
        title = "{A Study of Compact Object Mergers as Short Gamma-Ray Burst Progenitors}",
      journal = {\apj},
         year = 2006,
        month = sep,
       volume = {648},
       number = {2},
        pages = {1110-1116},
          doi = {10.1086/505169},
archivePrefix = {arXiv},
       eprint = {astro-ph/0601458},
 primaryClass = {astro-ph},
       adsurl = {https://ui.adsabs.harvard.edu/abs/2006ApJ...648.1110B}
}

@ARTICLE{Dominik2012,
       author = {{Dominik}, Michal and {Belczynski}, Krzysztof and {Fryer}, Christopher and {Holz}, Daniel E. and {Berti}, Emanuele and {Bulik}, Tomasz and {Mandel}, Ilya and {O'Shaughnessy}, Richard},
        title = "{Double Compact Objects. I. The Significance of the Common Envelope on Merger Rates}",
      journal = {\apj},
         year = 2012,
        month = nov,
       volume = {759},
       number = {1},
          eid = {52},
        pages = {52},
          doi = {10.1088/0004-637X/759/1/52},
archivePrefix = {arXiv},
       eprint = {1202.4901},
 primaryClass = {astro-ph.HE},
       adsurl = {https://ui.adsabs.harvard.edu/abs/2012ApJ...759...52D}
}

@ARTICLE{Gehrels04,
       author = {{Gehrels}, N. and {Chincarini}, G. and {Giommi}, P. and {Mason}, K.~O. and {Nousek}, J.~A. and {Wells}, A.~A. and {White}, N.~E. and {Barthelmy}, S.~D. and {Burrows}, D.~N. and {Cominsky}, L.~R. and {Hurley}, K.~C. and {Marshall}, F.~E. and {M{\'e}sz{\'a}ros}, P. and {Roming}, P.~W.~A. and {Angelini}, L. and {Barbier}, L.~M. and {Belloni}, T. and {Campana}, S. and {Caraveo}, P.~A. and {Chester}, M.~M. and {Citterio}, O. and {Cline}, T.~L. and {Cropper}, M.~S. and {Cummings}, J.~R. and {Dean}, A.~J. and {Feigelson}, E.~D. and {Fenimore}, E.~E. and {Frail}, D.~A. and {Fruchter}, A.~S. and {Garmire}, G.~P. and {Gendreau}, K. and {Ghisellini}, G. and {Greiner}, J. and {Hill}, J.~E. and {Hunsberger}, S.~D. and {Krimm}, H.~A. and {Kulkarni}, S.~R. and {Kumar}, P. and {Lebrun}, F. and {Lloyd-Ronning}, N.~M. and {Markwardt}, C.~B. and {Mattson}, B.~J. and {Mushotzky}, R.~F. and {Norris}, J.~P. and {Osborne}, J. and {Paczynski}, B. and {Palmer}, D.~M. and {Park}, H.-S. and {Parsons}, A.~M. and {Paul}, J. and {Rees}, M.~J. and {Reynolds}, C.~S. and {Rhoads}, J.~E. and {Sasseen}, T.~P. and {Schaefer}, B.~E. and {Short}, A.~T. and {Smale}, A.~P. and {Smith}, I.~A. and {Stella}, L. and {Tagliaferri}, G. and {Takahashi}, T. and {Tashiro}, M. and {Townsley}, L.~K. and {Tueller}, J. and {Turner}, M.~J.~L. and {Vietri}, M. and {Voges}, W. and {Ward}, M.~J. and {Willingale}, R. and {Zerbi}, F.~M. and {Zhang}, W.~W.},
        title = "{The Swift Gamma-Ray Burst Mission}",
      journal = {\apj},
         year = 2004,
        month = aug,
       volume = {611},
       number = {2},
        pages = {1005-1020},
          doi = {10.1086/422091},
archivePrefix = {arXiv},
       eprint = {astro-ph/0405233},
 primaryClass = {astro-ph},
       adsurl = {https://ui.adsabs.harvard.edu/abs/2004ApJ...611.1005G}
}

@ARTICLE{OConnor2022,
       author = {{O'Connor}, B. and {Troja}, E. and {Dichiara}, S. and {Beniamini}, P. and {Cenko}, S.~B. and {Kouveliotou}, C. and {Gonz{\'a}lez}, J.~B. and {Durbak}, J. and {Gatkine}, P. and {Kutyrev}, A. and {Sakamoto}, T. and {S{\'a}nchez-Ram{\'\i}rez}, R. and {Veilleux}, S.},
        title = "{A deep survey of short GRB host galaxies over z 0-2: implications for offsets, redshifts, and environments}",
      journal = {\mnras},
         year = 2022,
        month = oct,
       volume = {515},
       number = {4},
        pages = {4890-4928},
          doi = {10.1093/mnras/stac1982},
archivePrefix = {arXiv},
       eprint = {2204.09059},
 primaryClass = {astro-ph.HE},
       adsurl = {https://ui.adsabs.harvard.edu/abs/2022MNRAS.515.4890O}
}

@ARTICLE{Berger2010,
       author = {{Berger}, E.},
        title = "{A Short Gamma-ray Burst ``No-host'' Problem? Investigating Large Progenitor Offsets for Short GRBs with Optical Afterglows}",
      journal = {\apj},
         year = 2010,
        month = oct,
       volume = {722},
       number = {2},
        pages = {1946-1961},
          doi = {10.1088/0004-637X/722/2/1946},
archivePrefix = {arXiv},
       eprint = {1007.0003},
 primaryClass = {astro-ph.HE},
       adsurl = {https://ui.adsabs.harvard.edu/abs/2010ApJ...722.1946B}
}

@ARTICLE{1991AcA....41..257P,
       author = {{Paczynski}, Bohdan},
        title = "{Cosmological gamma-ray bursts.}",
      journal = {\actaa},
         year = 1991,
        month = jan,
       volume = {41},
        pages = {257-267},
       adsurl = {https://ui.adsabs.harvard.edu/abs/1991AcA....41..257P}
}
\bibliographystyle{aasjournalv7}

\appendix
\begin{figure}[!b]
    \centering
    \includegraphics[width=0.37\linewidth]{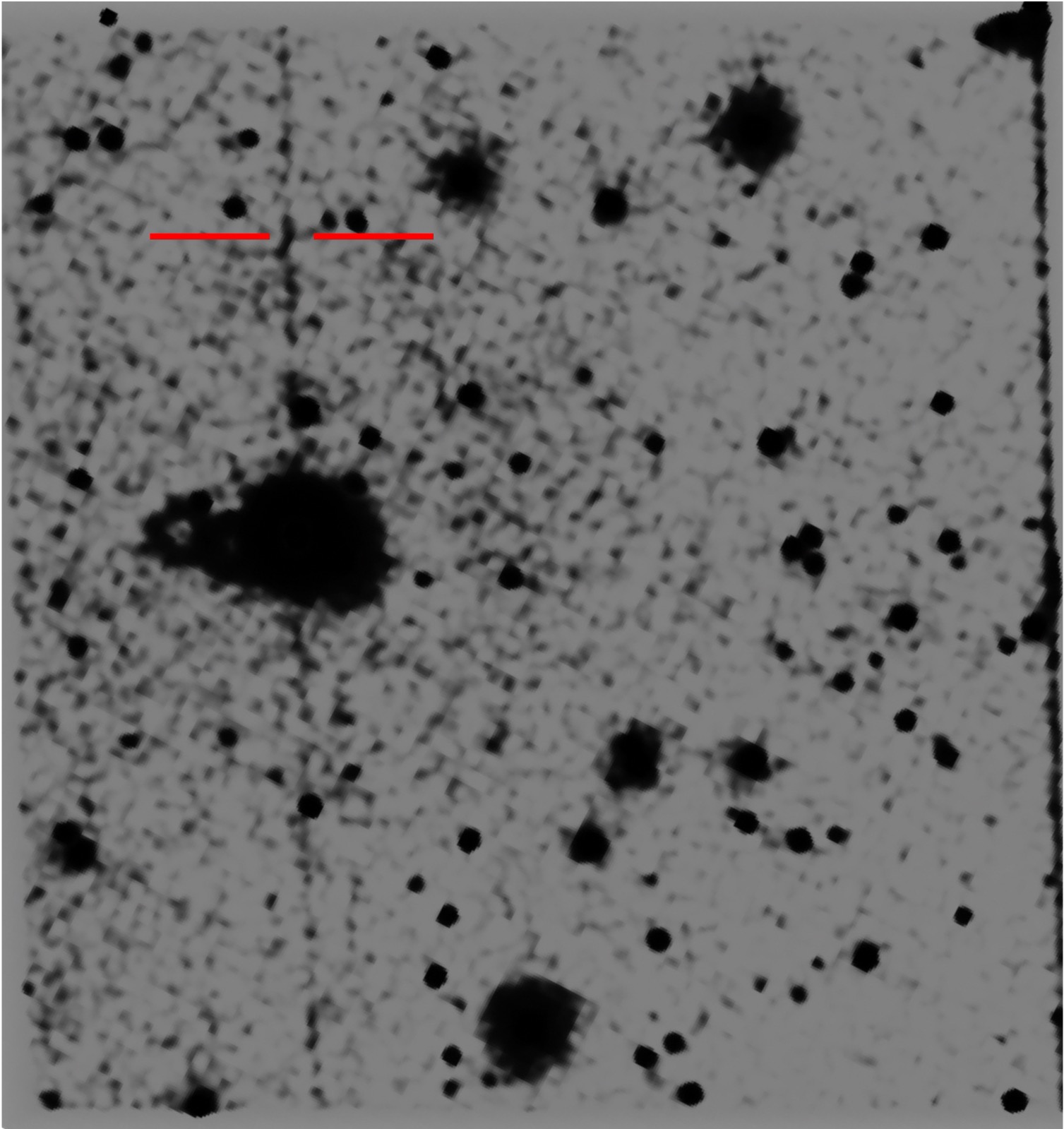}\hspace{0.3cm}
        \includegraphics[width=0.39\linewidth]{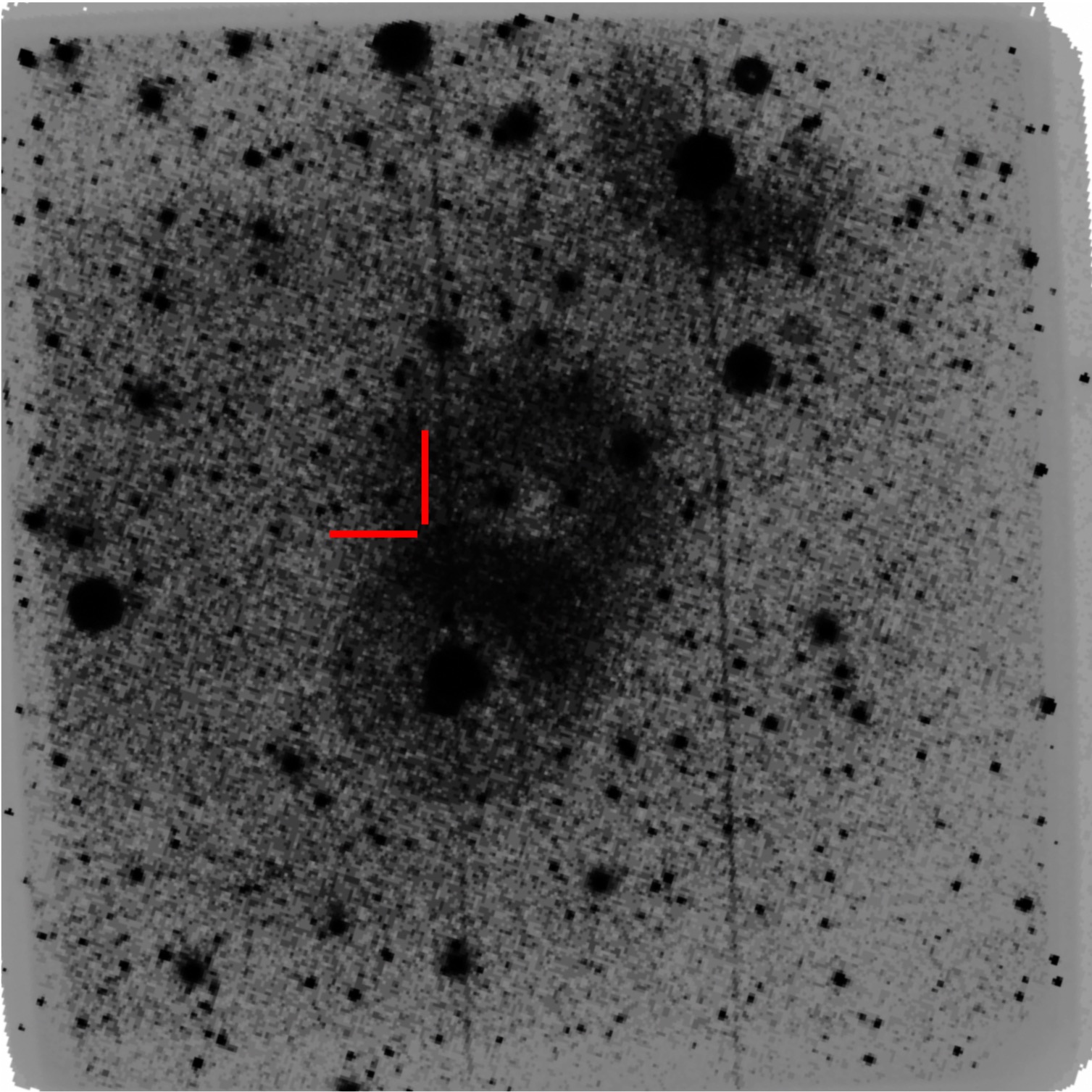}
    \caption{\textit{Left:} Optical image of the field, obtained by stacking the UVOT $u$, $b$, $v$ observations in segment 000. The red lines mark the GRB position affected by a read-out streak from the nearby star. \textit{Right:} UV image of the field, obtained by stacking the UVOT $w1$ observations in segment 001. The red lines mark the GRB position, falling within rings of scattered light. The colormaps are scaled using the histogram equalization function in ds9.}
    \label{fig:uvot}
\end{figure}
\section{VLT spectrum}
\label{appendix:vlt}

VLT/FORS2 spectroscopy of the GRB afterglow, presented in \citet{Stratta2007}, sets a direct constraint on the GRB distance scale. We therefore reanalyzed these observations to search for redshift-dependent spectral features.

The observations were carried out on 2006 December 2, starting at 01:18:55 UT ($T_0+0.4$\,d) with $2\times1800$\,s exposure time, using a 1\arcsec\ wide slit. 
The instrumental setup consisted of the 300V grism combined with the GG375 filter, covering approximately 3800--7600\,\AA.
After cosmic-ray removal from the raw frames, the archival data were reduced with the FORS spectroscopic pipeline through \texttt{EsoReflex} \citep{Freudling2013}. 
The reduced, stacked 2D spectrum reveals a faint trace at the GRB position, visible down to $\approx5100$\,\AA. At shorter wavelengths, the background noise prevents a secure identification of the source. Although no clear absorption or emission features are visible, the detection of the afterglow continuum sets an upper limit on the Lyman-limit wavelength and implies $z \lesssim (5100/912 - 1) \simeq 4.6$.

\section{UVOT images}\label{appendix:uvot}
The GRB position lies approximately 2\arcmin\ from  a bright ($V$\,$\simeq$\,11 mag) star, which affects the local background.  Thus, the limiting uncertainty is not only set by the Poisson noise but also by the accuracy with which the stellar contamination can be estimated.

Figure~\ref{fig:uvot} (left panel) shows the stack of the optical $u$, $b$, and $v$ images taken during the first epoch (segment 000). All images were reprocessed using the standard HEASoft \textit{Swift} tools and the relevant calibration files, corrected for modulo-8 fixed-pattern noise, astrometrically aligned using \texttt{uvotskycorr}, then co-added with \texttt{uvotimsum}. 
The optical stack shows that a read-out streak from the nearby bright source crosses the afterglow position. 
This artifact also affects images taken with the $wh$ filter, for which
we report a detection in Table~\ref{tab:photometry}. 
However, our conclusions do not depend on this detection, but rather on the faintness of the optical counterpart at blue wavelengths. Thus, they would remain unchanged even if the source was treated as an upper limit.

Figure~\ref{fig:uvot} (right panel) shows the stack of the UV $w1$ images taken during the second epoch (segment 001). The location of the streak has moved. However, scattered-light rings from nearby bright stars contaminate the GRB position.
These rings produce a higher and spatially variable background, which might affect the detection significance and photometry of faint sources. Our re-analysis does not reproduce the marginal $w1$ detection reported by \citet{Mao2026}. 

\section{Chance coincidence probability}\label{appendix:pcc}

When a direct measurement of the distance scale is not available, a GRB is commonly associated with the
host galaxy that has the lowest chance coincidence probability $P_{\rm cc}$. This quantity estimates the likelihood of finding, by chance, a galaxy of comparable or greater brightness within a given angular distance from the GRB position \citep{Bloom2002}.
Low values of $P_{\rm cc}$ therefore indicate that the GRB-galaxy alignment is unlikely to be random.

In Table~\ref{tab:pcc_candidates}, we report the brightness, projected angular offsets, and resulting $P_{\rm cc}$ values for the three candidate host galaxies G1, G2, and G3. For comparison, we also report the values previously derived in the literature by \citet{Fong2013} and \citet{Mao2026}. 

We compute \(
P_{\rm cc}=1-\exp\left[-\pi r_{\rm eff}^{2}\,\sigma(\le m)\right]
\)
where \(\sigma(\le m)\) is the sky density of sources brighter than magnitude \(m\) derived from the galaxy number counts of \citet{Hogg1997}
for the optical bands and of \citet{Windhorst2023} for the nIR bands.
The effective radius $r_{\rm eff}$ is defined following \citet{Bloom2002}. For G1 and G2 $r_{\rm eff}$ is dominated by the large angular offset, whereas for G3 we find $r_{\rm eff}\approx$0.4\arcsec.
In agreement with \citet{Mao2026}, we find that G1 has the lowest $P_{\rm cc}$, ranging from 9\% to 11\% depending on the adopted filter. The galaxy G2 is characterized by a higher probability, ranging from 15\% to 22\%. 
The galaxy G3 is not detected in the standard $R$ and $F160W$ filters, and those bands provide only lower limits on its $P_{\rm cc}$. 

To estimate the chance coincidence probability for G3, we calculate the probability of intercepting a source 28.5 mag or brighter  in our $F322W2$ image. We use the segmentation map produced by Source Extractor \citep{Bertin1996}, reject stellar sources with \texttt{CLASS\_STAR}$\geq$ 0.3, and manually mask residual saturated stars and diffraction spikes. In the cleaned segmentation map, sources with \texttt{MAG\_AUTO}\,$\lesssim 28.5$ AB mag occupy $\simeq 11\%$ of the total image area. The probability that a random position lies within $0.4\arcsec$ of any of these sources is $P_{cc}\simeq 13\%$.
As an alternative method, we treat the observed $F150W2$ and $F322W2$ magnitudes as approximate values for the $F150W$ and $F356W$ bands, respectively, and compare them with the near-infrared galaxy number counts of \citet{Windhorst2023}. 
We obtain values of $P_{\rm cc}$ ranging from 13\% to 5\%, comparable to the segmentation map method. 

\begin{table}
\centering
\caption{Chance coincidence probabilities for the three candidate host galaxies of GRB061201.
The table lists the angular offset of each galaxy's center from the
afterglow position (from \citealt{Fong2013} and this work), 
the measured optical (\(R\)-band; this work) and near-infrared magnitude (\(F160W\); \citealt{Fong2013}), and the
corresponding probability of chance coincidence, \(P_{\rm cc}\), inferred from
the relevant galaxy number counts. For G3, we report limits from those bands and
use the \(F322W2\) image to estimate its \(P_{\rm cc}\) as described in the text. 
The final two columns list the values of \(P_{\rm cc}\) reported in the literature
(F13: \citealt{Fong2013}, and M16: \citealt{Mao2026}). Magnitudes are in the AB
system, and offsets are given in arcseconds.
}
\label{tab:pcc_candidates}
\begin{tabular}{lccccccccc}
\hline
Galaxy & Offset  & $R$  & $P_{\rm cc}(r)$ 
       & $F160W$  &  $P_{\rm cc}(F160W)$ 
       & $F322W2$ & $P_{\rm cc}({\rm F322W2})$  & $P_{\rm cc}$ & $P_{\rm cc}$ \\
& (arcsec) & (AB) &  & (AB) & & (AB) &  & FB13 & M16\\
\hline
G1 & 16.25\,$\pm$\,0.03 & $19.42 \pm 0.11$ & 0.11    & $18.63 \pm 0.01$ & 0.10 & -- & -- & 0.07 &  0.11 \\
G2 &  1.80\,$\pm$\,0.03 & $26.0 \pm 0.2$   & 0.22    & $24.46 \pm 0.03$ & 0.15  & -- & -- & 0.07 &  0.18 \\
G3 &  0.21\,$\pm$\,0.13 & $<$26.6          & $>$0.02 & $<$26.4          & $>$0.01 & $28.42 \pm 0.15$ & 0.13 & --   &  0.46 \\
   &                    & & & $F150W$\,$\approx$\,29.8  & $\approx$\,0.13 & $F356W$\,$\approx$\,28.4 & $\approx$\,0.05 &    &  \\

\hline
\end{tabular}
\end{table}

\section{Broadband SED with an SMC extinction law}\label{appendix:smc}

The typical extinction law of short GRB sight lines is not well constrained.  Therefore, in addition to the MW extinction curve, we repeated the SED modeling adopting a Small Magellanic Cloud (SMC) dust law for the host galaxy \citep{Pei1992}. For this case, we converted reddening to visual extinction using $R_V = 2.93$.

\begin{figure}
    \centering
    \includegraphics[width=0.45\linewidth]{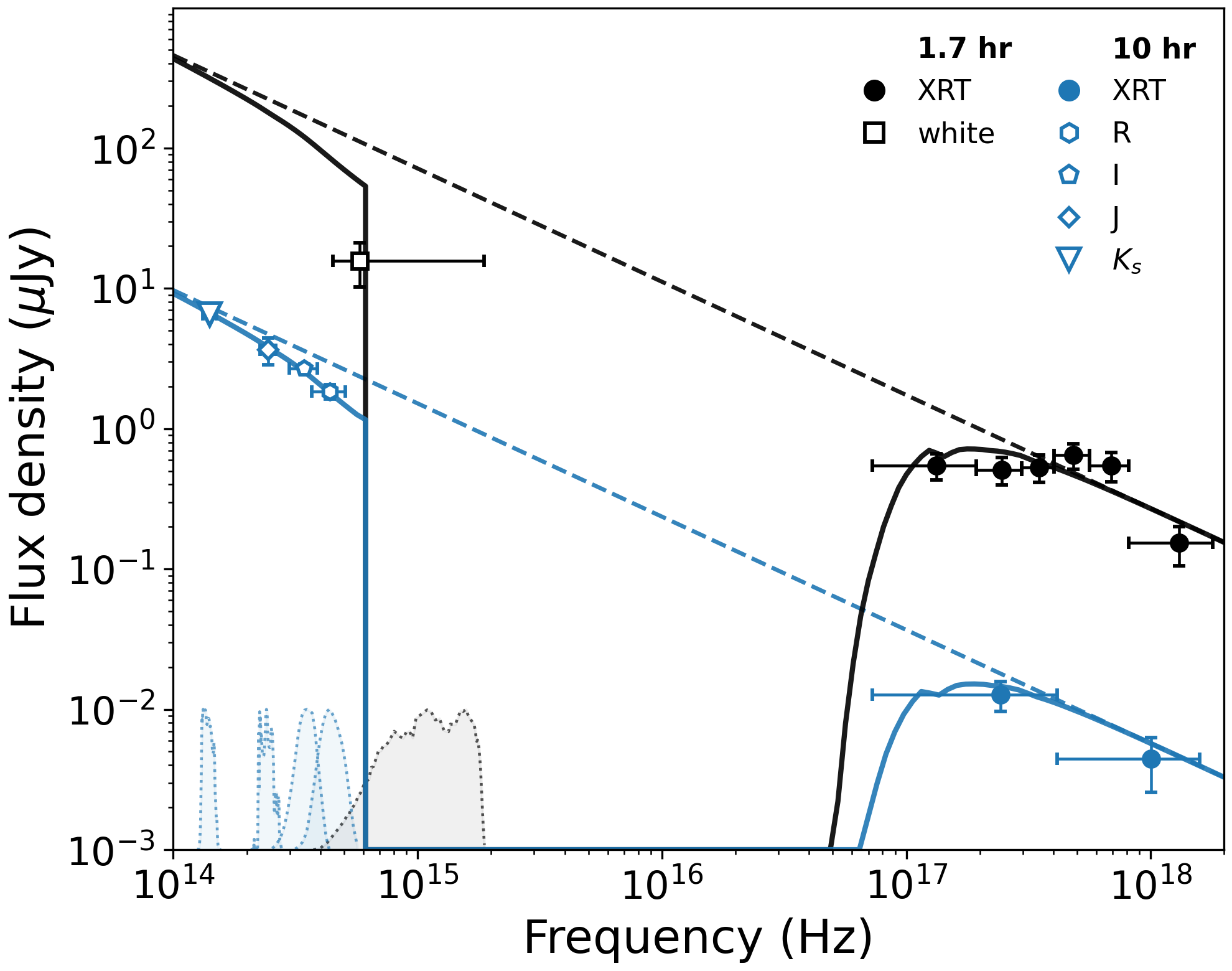}\hspace{0.2cm}
    \includegraphics[width=0.45\linewidth]{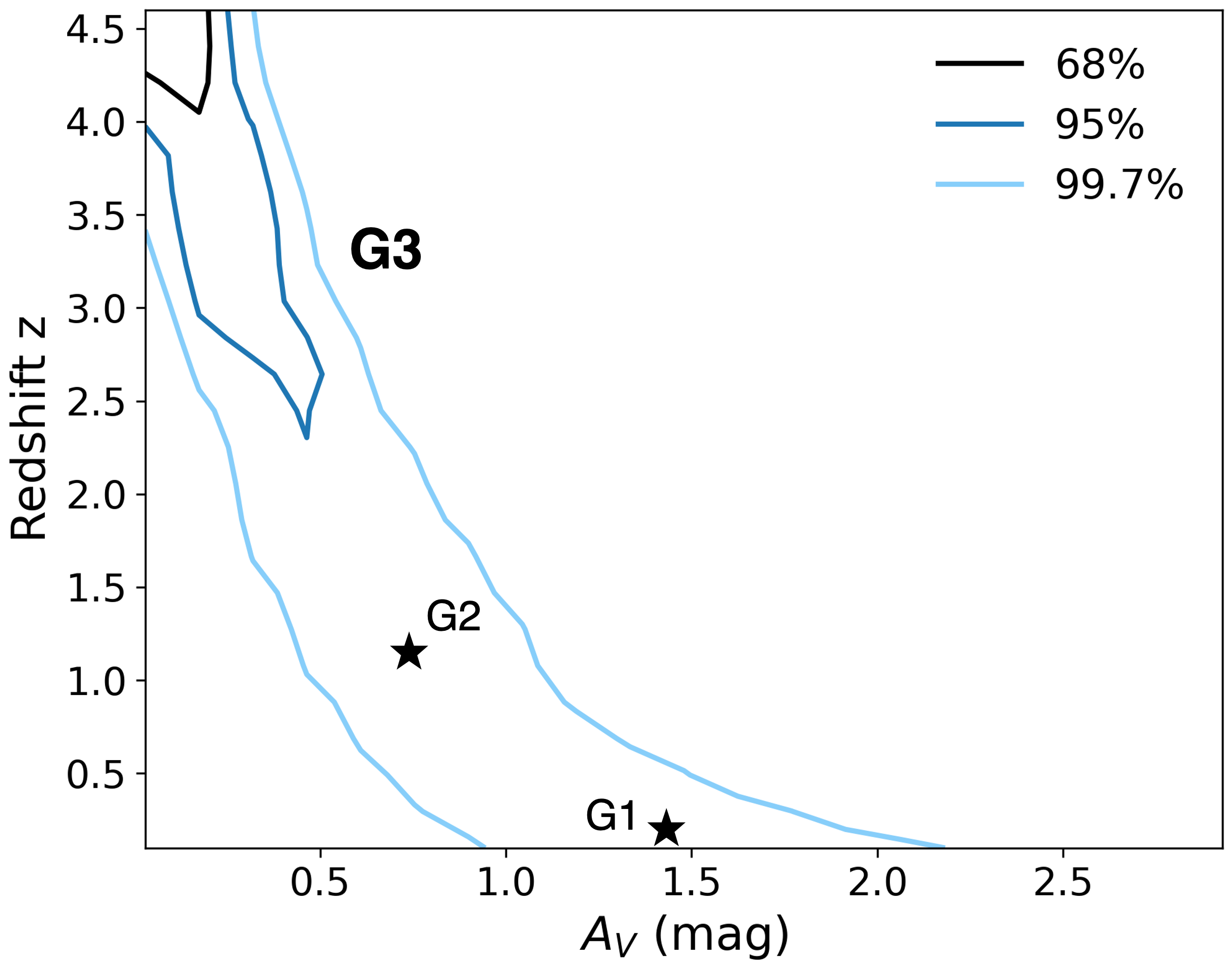}
    \caption{Confidence contours for intrinsic extinction $A_V$ and redshift $z$, assuming an SMC extinction law. Contours show the 68\%, 95\%, and 99.7\% confidence regions. The favorite model corresponds to a high-redshift origin, consistent with an association with G3. The stars indicate the best fit values for G1 and G2.}
    \label{fig:sedSMC}
\end{figure}

Our best fit model, shown in Figure~\ref{fig:sedSMC} (left panel), is described by a redshift $z$\,=\,4.4$^{+0.2}_{-0.6}$, a photon index \(\Gamma = 1.80 \pm 0.04\), an intrinsic hydrogen column density
\(N_{\rm H,z} = (1.6^{+1.6}_{-1.3})\times10^{22}\,{\rm cm^{-2}}\), and 
reddening \(E(B-V)=0.03\pm0.02\). The Galactic foreground values were kept fixed at \(E(B-V)=0.065\) and \(N_{\rm H}=6.7\times10^{20}\,{\rm cm^{-2}}\) \citep{SF2011,Willingale2013}.

As in the MW case, the optical flux can be reproduced by a degeneracy between dust extinction and redshift. The confidence contours in the $A_V$--$z$ plane show an elongated shape (Figure~\ref{fig:sedSMC}, right panel), reflecting the trade-off between these two effects: lower-redshift solutions require larger rest-frame extinction, whereas higher-redshift solutions require less dust because absorption by neutral hydrogen suppresses the observed flux. 
The resulting contours are smoother than the MW contours because the SMC curve is featureless through the rest-frame UV compared with the MW law. In particular, it lacks the pronounced 2175 \AA\ extinction bump. 

\label{lastpage}

\end{document}